\documentclass[aps,prl,reprint,superscriptaddress,letterpaper,twocolumn,floatfix]{revtex4-2}
\usepackage{amsmath,amssymb,mathtools,dsfont,textgreek,soul}

\usepackage{graphicx,setspace}
\usepackage[usenames, dvipsnames, x11names]{xcolor}
\usepackage[colorlinks]{hyperref}
\usepackage{physics}
\usepackage[capitalise]{cleveref}

\begin{document}

\title{Floquet Reservoir Engineering for Remote Logical Entanglement}

\author{Mingxing Yao}
\email{mxyao@uchicago.edu}
\affiliation{Pritzker School of Molecular Engineering, University of Chicago, Chicago, IL 60637, USA}

\author{Aashish A. Clerk}
\affiliation{Pritzker School of Molecular Engineering, University of Chicago, Chicago, IL 60637, USA}

\begin{abstract}

Implementing controlled dissipative dynamics is a powerful approach for state preparation in a variety of contexts, including the preparation of remote entangled states.  Here, we show that by going beyond the standard setting of time-independent dissipative dynamics, one can realize even more powerful non-unitary protocols.  We introduce dissipative Floquet protocols for stabilizing remote entanglement of logical qubits,  where continuously-running dissipation is interleaved with a periodic sequence of unitary gates.  These protocols harness existing experimental capabilities, and overcome time-entanglement limits that constrain standard approaches.  They also implement an autonomous form of entanglement distillation.  We show how these protocols give enhanced protection against waveguide loss, and as an example, analyze a specific implementation using cat-qubits and transmons in a superconducting circuit. 

\end{abstract}

\maketitle

{\it Introduction.}---  A powerful and somewhat counterintuitive method for quantum state preparation is reservoir engineering, the use of tailored dissipative dynamics to stabilize a target state or manifold \cite{Poyatos-physRevLett-1996, Diehl-naturePhysics-2008, Verstraete-naturePhysics-2009-y,Plenio-physRevA-1999-c,Kraus-physRevA-2008-d}.  It has been implemented in a variety of experimental platforms (see e.g.~\cite{Barreiro-nature-2011-o,Shankar-nature-2013,Lin-nature-2013,Krauter-physRevLett-2011-x,Ma-nature-2019-i,Kienzler-science-2015-m,Leghtas-science-2015-g,Lescanne-natPhys-2020-w,Murch-physRevLett-2012-a}), including demonstrations of autonomous quantum error correction~\cite{Gertler-nature-2021-e,Lachance-Quirion-physRevLett-2024-p,Li-natCommun-2024-k}.  A particularly compelling application is the dissipative stabilization of remote entanglement (via waveguides or transmission lines) \cite{Gonzalez-Tudela-physRevLett-2011-i,Pichler-physRevA-2015-l,Lingenfelter-physRevX-2024, Stannigel-newJPhys-2012-i,Motzoi-physRevA-2016-t,Irfan-arxivquant-ph-2025-w, Andres-Juanes-arxivquant-ph-2025-f, Irfan-physRevRes-2024-v,Shah-prxQuantum-2024-m,Almanakly-arxivquant-ph-2026-d,Brown-natCommun-2022-h,Ramos-physRevLett-2014-r,Didier-physRevAcollPark-2018-a,Muschik-physRevA-2011-g}, a resource crucial to quantum networking and distributed computing~\cite{Kimble-nature-2008, Knorzer-arxivquant-ph-2025-m}.  

Standard reservoir engineering employs time-independent dissipative dynamics.  While powerful, it has known limitations. For example, most protocols for dissipative remote entanglement suffer from a fundamental time-entanglement tradeoff that precludes generating maximally-entangled states~\cite{Pocklington-prxQuantum-2024-w,Brown-natCommun-2022-h}. Can one transcend these and other limitations by considering a more general kind of tailored dissipative dynamics?  In this paper, we show that this is indeed true.  We consider a very general Floquet reservoir engineering protocol for remote entanglement, where time-independent dissipation is interleaved with a periodic sequence of fast unitary gates, resulting in time-periodic non-unitary dynamics. We introduce a class of new protocols that avoid the above time-entanglement tradeoff, and also enable a crucial new functionality: the ability to dissipatively stabilize entanglement between remote protected or logical qubits that are not directly coupled to the engineered dissipative reservoir. Our work complements studies of repeated-channel dynamics in other settings (e.g.~in collision models~\cite{Ciccarello-physRep-2022-g,Verstraete-naturePhysics-2009-y,Karevski-physRevLett-2009-b,Schindler-natPhys-2013-e,Barreiro-nature-2011-o,Mi-science-2024-u}). 

%%%%%%%%%%%%%%%%%%%%%%%%%%%%%
\begin{figure}[th]
  \includegraphics{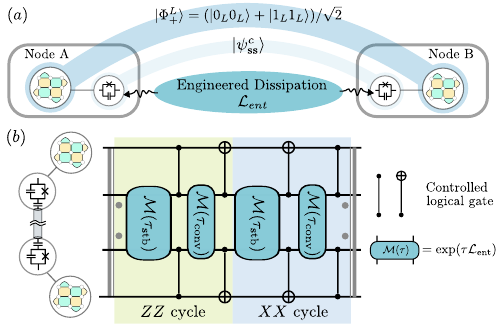}
  \caption{\textbf{Floquet reservoir engineering to stabilize remote logical Bell states}
  (a) Two remote computational nodes (A and B) each consist of a logical qubit (depicted as a surface code patch) and a communication qubit (represented by a transmon). The communication qubits are coupled via an engineered entangling dissipative bath $\mathcal{L}_{\rm ent}$. (b) Circuit description of one period of the protocol: entangling engineered dissipation acts continuously on the communication qubits, interspersed with fast controlled unitary gates.  The gates partition the dissipation into finite-time channels $\mathcal{M}(\tau) \equiv \exp(\tau \mathcal{L}_{\rm ent})$.}
  \label{fig-1:Schematic}
\end{figure}
%%%%%%%%%%%%%%%%%%%%%%%%%%%%%

Our protocol is sketched in \cref{fig-1:Schematic}, and involves two remote nodes (each with a logical and a communication qubit), coupled to a common, extended dissipative reservoir (a waveguide or transmission line).  The protocol uses two generic ingredients that have been independently realized in many platforms:  (1) a time-independent engineered reservoir that stabilizes non-maximal entanglement between a pair of remote communication qubits (see e.g.~\cite{Irfan-arxivquant-ph-2025-w,Andres-Juanes-arxivquant-ph-2025-f,Shah-prxQuantum-2024-m}), and (2) the ability to perform fast CNOT or CZ unitary gates between the communication qubits and the protected logical qubits at each node 
\cite{Guillaud-physRevX-2019-a,Puri-sciAdv-2020-b,Reinhold-natPhys-2020-r,Heeres-natCommun-2017-l}.
By periodically interleaving these instantaneous gates with continuous dissipation, our protocol deterministically and autonomously stabilizes in finite time a {\it maximally} entangled state between two remote protected qubits. It functions like an autonomous entanglement distillation engine, concentrating the weak bath-induced entanglement of the communication qubits into a near-maximal entangled state of the protected qubits.  We stress that our protocol does not require any explicit measurements or classical communication between nodes.  As we discuss below, it also offers advantages (both practical and fundamental) over previous proposals for autonomous entanglement distillation~\cite{Didier-physRevAcollPark-2018-a,Vollbrecht-physRevLett-2011-y}.   

We apply our general construction to two paradigmatic choices of entangling reservoirs. These combine 
collective dissipation generated by a waveguide with effective driving, either in the form of two-mode squeezed light \cite{Kraus-physRevLett-2004-g,Agusti-arxivquant-ph-2022-t,Andres-Juanes-arxivquant-ph-2025-f,Govia-physRevRes-2022-d}, or simple Rabi driving \cite{Stannigel-newJPhys-2012-i,Irfan-arxivquant-ph-2025-w}.  We identify optimal parameter regimes, and demonstrate that even with loss, large amounts of logical entanglement can be stabilized, amounts that surpass what is possible with communication qubits alone.  We also analyze a concrete, experimentally-relevant example based on logical cat qubits and transmon communication qubits in a superconducting modular processor, showing the utility of our approach in a noise-bias preserving framework.

%%%%%%%%%%%%%%%%%%%%%%%%%%%%

{\it Model and protocol.}--- Consider two spatially remote computational nodes, labeled $A$ and $B$, each maintaining a protected (logical) qubit and an auxiliary physical communication qubit [\cref{fig-1:Schematic}(a)]. Our goal is to stabilize the logical Bell state $\ket{\Phi^L_+} = (\ket{0^L_A 0^L_B} + \ket{1^L_A 1^L_B})/\sqrt{2}$, uniquely fixed by the stabilizer expectation values $\langle \hat X^L_A \hat X^L_B \rangle = \langle \hat Z^L_A \hat Z^L_B \rangle = 1$. Here $\ket{0^L}/\ket{1^L}$ denotes the logical qubit basis and $\hat X^L$ , $\hat Y^L$ and $\hat Z^L$ are the logical Pauli operators. The non-local resource linking the nodes is a waveguide or transmission line that couples only to the communication qubits.  Combined with local Hamiltonian terms, it generates an entangling time-independent dissipation on the communication qubits, described by a Lindblad-form master equation~\cite{Lindblad-communMathPhys-1976,Gorini-jMathPhys-1976}:
\begin{equation}
    \partial_t \hat \rho = - i[\hat H^c_{AB} ,\hat \rho] + \gamma \sum_j \mathcal{D}[\hat L^c_j] \hat \rho
    \equiv \mathcal{L}_{\rm ent} \hat \rho,   \label{eq:Lindbladian}
\end{equation}

where $\hat H^c_{AB}$ is a Hamiltonian, $\hat L^c_j$ are jump operators, all acting on the communication qubits.  We define $\mathcal{D}[\hat A] \hat \rho = \hat A \hat \rho \hat A^\dagger - \frac{1}{2}\left\{ \hat A^\dagger \hat A, \hat \rho\right\}$.  We require that this dynamics yield a unique, pure entangled steady state:
\begin{equation}
    \mathcal{L}_{\rm ent} \left(\ket{\psi^c_{ss}} \bra{\psi^c_{ss}} \right) = 0, 
    \,\,\,
    \ket{\psi^c_{ss}} = u\ket{0^c_A 0^c_B} + v\ket{1^c_A 1^c_B}. 
    \label{eq:dark-state}
\end{equation}
Here $\ket{0^c},\ket{1^c}$ ($u,v$) are the Schmidt basis (coefficients), and $\ket{g^c}/\ket{e^c}$ denote computational basis states.
{We also have the mild requirement that the even-parity subspace spanned by $\{\ket{0_A^c 0_B^c},\ket{1_A^c 1_B^c} \} $ is not closed under the dynamics. } 
As a concrete example, consider the distributed two-mode squeezing scheme of~\cite{Kraus-physRevLett-2004-g,Govia-physRevRes-2022-d}. In this case:
\begin{equation}
\begin{split}
  \mathcal{L}_{\text{TMS}}\rho = \gamma\, &\mathcal{D}[\cosh r\,\hat\sigma_A^- - \sinh r\,\hat\sigma_B^+]\rho \\
  &\quad + \gamma\, \mathcal{D}[\cosh r\,\hat\sigma_B^- - \sinh r\,\hat\sigma_A^+]\rho.
\end{split}
\label{eq:TMSLindblad}
\end{equation}

To make this time-independent dissipative dynamics into a Floquet reservoir engineering scheme that entangles the logical qubits, we now add a periodic sequence of node-local unitary gates.  
During each protocol period depicted in~\cref{fig-1:Schematic}(b), fast local unitary gates, effectively instantaneous on the dissipative timescale, separate this continuous dissipation into finite segments, each described by the channel $\mathcal{M}(\tau) \equiv e^{\mathcal{L}_{\rm ent}\tau}$. A single period consists of two subroutines $\mathcal{E} = \mathcal{E}_{ZZ} \circ \mathcal{E}_{XX}$, each having two controlled unitary gates.
As we will show below $\mathcal{E}_{ZZ}$ ($\mathcal{E}_{XX}$) acts to stabilize $ZZ$ ($XX$) logical correlations.  For $\mathcal{E}_{ZZ}$ we use two controlled unitary gates:
\begin{equation}
\mathcal{E}_{ZZ} \;=\; \mathcal{C}_X\,\circ\,\bigl[\mathbb{I}_L\otimes\mathcal{M}(\tau_{\rm conv})\bigr]\,\circ\,\mathcal{C}_Z\,\circ\,\bigl[\mathbb{I}_L\otimes\mathcal{M}(\tau_{\rm stb})\bigr],
\label{eq:composite-channel}
\end{equation}
where $\mathcal{C}_{X(Z)}$ are local control-$X^L(Z^L)$ gates, with the communication qubit as control and the logical qubit as target. The dissipation-only evolution times $\tau_{\rm stb}$ and $\tau_{\rm conv}$ are set by the timing of the unitaries.   
$\mathcal{E}_{XX}$ is constructed in an analogous fashion, with the order of unitary gates interchanged (see \cref{fig-1:Schematic}b).  

{\it Stabilization mechanism.}-- Surprisingly, no matter what the choice of time intervals $\tau_{\rm stb}$,$\tau_{\rm conv}$, as long as the bath entanglement parameter $|u|,|v| > 0$, the composite channel $\mathcal{E}$ has a unique pure steady state with maximal logical-qubit entanglement \footnote{While one can easily check analytically that this is a steady state of the channel, we do not have an analytic proof of uniqueness for arbitrary parameter regime; we can analytically confirm the convergence for the case $\tau_{\rm stb} \gg \tau_{\rm ent}$ with~\cref{eq:pconvDefinition}. Our conclusion relies on numerics beyond this scope.}:
\begin{equation}
  \mathcal{E} (\ket*{\Psi^{\rm tot}_{ss}} \bra*{\Psi^{\rm tot}_{ss}}) = \ket*{\Psi^{\rm tot}_{ss}} \bra*{\Psi^{\rm tot}_{ss}} , 
  \hspace{0.20 cm}
  \ket*{\Psi^{\rm tot}_{ss}} = \ket*{\Phi^L_+}\otimes\ket{\psi^c_{\rm ss}}.
\end{equation}
At a heuristic level, each application of the channel $\mathcal{E}$ serves to both maintain the communication qubits in the entangled state $\ket{\psi^{c}_{ss}}$ of Eq.~\eqref{eq:dark-state}, and also distill some of this entanglement into the two logical qubits.  
Intuition into reservoir-engineering protocols is often best established by viewing them as autonomous versions of measurement-plus-feedback protocols.  This also applies to our scheme: it may be viewed as an autonomous analogue of protocols that generate remote entanglement via explicit parity measurements and feedback~\cite{Riste-nature-2013, Roch-physRevLett-2014, Andersen-npjQuantumInf-2019-a}. We stress however that our protocol has no actual measurements or classical communication.  
To understand the above connection, we start with the $\mathcal{E}_{ZZ}$ subroutine, and explain heuristically what each of the four steps (two dissipative evolutions, two unitaries) accomplish.  

(i) {\it Resource Preparation $\mathbb{I}_L\otimes\mathcal{M}(\tau_{\rm stb})$:} $\mathcal{E}_{ZZ}$ begins by applying dissipation to the communication qubits for a time $\tau_{\rm stb}$, while the logical qubits idle.  This step establishes entanglement in the communication qubits.  To simplify our analysis, we take $\tau_{\rm stb}$ longer than the characteristic relaxation time $\tau_{\rm ent}$ of $\mathcal{L}_{ent}$, such that the communication-qubits are (to a good approximation) in the state $\ket{\psi^c_{ss}}$ at the end of this step. 

(ii) {\it Parity Encoding $\mathcal{C}_Z$:} Next, a fast control-$Z^L$ gate is applied locally at each node (where $Z$ for the communication qubits is defined with respect to the Schmidt basis).  This gate serves to encode global $Z$ parity of the logical qubits onto the communication qubits: the latter evolve into an orthogonal state $\ket*{\psi_{\rm det}^c}$ if the logical state has odd parity.  To be explicit, 
suppose the logical qubits start in a pure state which we decompose into parity even and odd components: 
$| \psi^L \rangle = \ket*{\psi_{e}^L} + \ket*{\psi_{o}^L}$, with $\hat Z^L_A \hat Z^L_B \ket*{\psi_{e/o}^L} = \pm \ket*{\psi_{e/o}^L}$.  Then step (ii) performs the unitary mapping:
\begin{equation}
    | \psi^L \rangle \otimes |\psi^c_{ss} \rangle  \rightarrow 
    \ket*{\psi^L_{e}}\otimes \ket*{\psi^c_{ss}} + \ket*{\psi_o^L} \otimes \ket*{\psi_{\rm det}^c}. \label{eq:parity-encoding}
\end{equation}
with $\ket*{\psi_{\rm det}^c} \equiv u \ket{0_A^c 0_B^c} - v \ket{1_A^c 1_B^c}$.

Note that while information on the logical parity is now available in the communication qubits, it is {\it only} encoded globally, and not accessible through local measurements at each node.

(iii) {\it Bath-enabled information transduction $\mathbb{I}_L\otimes\mathcal{M}(\tau_{\rm conv})$:} The third step again involves dissipative evolution of the communication qubits via $\mathcal{M}(\tau_{\rm conv})$ while the logical qubits idle. Unlike step (i), the goal here is {\it not} to reset the state of the communication qubits.  Instead, the aim is to transduce the logical parity information stored in the communication qubits into something that can be used locally at each node.  Looking at the the final state in Eq.~\eqref{eq:parity-encoding}, we note that if the communication qubits are in state $| \psi^c_{ss} \rangle$, then $\mathcal{M}(\tau)$ does nothing to them, whereas if they are in the state 
$| \psi^c_{\rm det} \rangle$, $\mathcal{M}(\tau)$ will cause evolution. 
As by assumption $\mathcal{L}_{\rm ent}$ does not preserve even parity, there will be evolution times $\tau > 0$ where the evolved state has a non-zero probability to overlap with odd parity subspace, i.e. $\exists \tau > 0$ such that
\begin{equation}
  p_{\text{conv}}(\tau) 
  \equiv \tr (\hat P_{\rm odd}^c \mathcal{M}(\tau) \left(\ket{\psi_{\rm det}^c}\bra{\psi_{\rm det}^c} \right)) > 0, 
  \label{eq:pconvDefinition}
\end{equation}
where $\hat P_{\rm odd}^c$ is the projector onto the odd $Z$-parity subspace.  
For a suitable $\tau = \tau_{\rm conv}$, we have that (with finite probability) logical parity information is encoded in the (computational-basis) $Z$ parity of the communication qubits, something that can be used locally.  %As we now show, this in turn can be used locally to ``correct" the state of the logical qubits.   
We thus see that $p_{\text{conv}}$ serves as a kind of detection probability.

(iv) {\it Feedback $\mathcal{C}_X$:} The last step in $\mathcal{E}_{ZZ}$ is a pair of $\mathcal{C}_X$ (controlled-$X^L$) gates (one at each node). This unitary effectively takes the logical parity information encoded in the communication qubits, and uses it to perform a ``correction" to the logical qubits.  
More concretely, if the evolution in step (iii) results in an odd parity communication qubit state, then a single local $X$ (either on $A$ or $B$) will be performed to the logical state.  This effectively converts an odd parity logical state to an even parity state, pushing the state towards the desired even-parity subspace where $\langle Z^L_A Z^L_B \rangle = 1$.  In contrast, if the communication qubits start in an even parity state, then there is no change in logical parity.  
This last feedback step serves to push the logical qubits towards the desired target entangled state.  

We now have a heuristic understanding of how the subroutine $\mathcal{E}_{ZZ}$ functions, and see how it mimics (in a fully autonomous fashion) a protocol based on parity measurements and feedback.  The subroutine $\mathcal{E}_{XX}$ can be understood in an analogous fashion.  The net result is a channel that pushes the logical state towards having even $Z$ and $X$ parities, which necessarily implies stabilization of 
$\ket{\Phi^L_+}$.  As one might expect, the fully autonomous nature of our protocol gives it several advantages over approaches that explicitly use measurement and feedback; we discuss this more in the End Matter. 

{\it Stabilization dynamics.}--
To get a sense of the speed of our protocol, we consider the evolution of logical correlators. 
In the limit $\tau_{\rm stb} \gg \tau_{\rm ent}$ (i.e.~each subroutine stabilizes the communication qubits to $\ket{\psi^c_{ss}}$), the single-period channel $\mathcal{E}$ maps them as (see \cite{supp}):
\begin{align}
  \langle Z_A^L Z_B^L \rangle_{n+1} &= \langle Z_A^L Z_B^L \rangle_n + p_{\text{conv}} (1 - \langle Z_A^L Z_B^L \rangle_n), \label{eq:zz-recursion}\\
  \langle X_A^L X_B^L \rangle_{n+1} &= \langle X_A^L X_B^L \rangle_n + p_{\text{conv}} (1 - \langle X_A^L X_B^L \rangle_n), \label{eq:xx-recursion}
\end{align}
with the correlator $\langle \hat O_A^L \hat O_B^L \rangle_n = \tr (\hat O_A^L \hat O_B^L \mathcal{E}^n (\hat \rho_0))$.  Here, $p_{\rm conv}$ is the probability that odd logical parity leads to an odd-parity communication-qubit state in step (iii), c.f.~Eq.~\eqref{eq:pconvDefinition}. 
Both logical stabilizers obey the same map and converge monotonically to unity, $1-\langle X_A^L X_B^L \rangle_n \sim (1-p_{\text{conv}})^n$. We define the dimensionless convergence rate $\Gamma$ via the decay of the state fidelity $F_n = \bra{\Psi^{\rm tot}_{ss}} \mathcal{E}^n( \hat \rho_0) \ket{\Psi^{\rm tot}_{ss}} \sim \exp(-\Gamma n)$.  As $F \geq \frac{\langle X^L_A X^L_B \rangle + \langle Z^L_A Z^L_B \rangle}{2}$, we have to a good approximation $\Gamma \simeq - \ln (1-p_{\text{conv}})$ (something that is verified by our full numerics).
We see that the speed of our protocol depends crucially on $p_{\text{conv}}$ (and hence $\tau_{\rm conv}$).  There will also be a dependence on $\tau_{\rm stb}$ (which was assumed above to be large), and the nature of the engineered Lindbladian $\mathcal{L}_{\text{ent}}$ (and entanglement parameter $v$).  

%%%%%%%%%%%%%%%%%%%%%%%%%%%%%
\begin{figure}[tb]
  \includegraphics{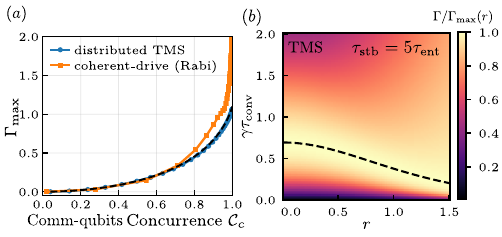}
  \caption{\textbf{Optimized rate for stabilizing logical Bell state.}
    (a) Maximized dimensionless logical stabilization rate $\Gamma_{\max}$ as a function of the communication qubit's steady-state $\ket{\psi^c_{\rm ss}}$ concurrence $\mathcal{C}_c[v]$, for both the TMS and Rabi-drive entangling reservoirs. Dashed line: analytic result for the TMS scheme. 
    (b) Normalized logical stabilization rate $\Gamma/\Gamma_{\max}(r)$ for the TMS scheme, as a function of the conversion time $\gamma\tau_{\rm conv}$ and the squeezing parameter $r$ at fixed preparation time $\tau_{\rm stb} = 5\tau_{\rm ent}$. The dashed curve marks the analytical optimal $\gamma\tau_{\rm conv}^*$.}
  \label{fig-2:numerics}
\end{figure}
%%%%%%%%%%%%%%%%%%%%%%%%%%%%%

We next analyze our general protocol for two concrete examples for $\mathcal{L}_{\rm ent}$
\footnote{Note that both of these examples satisfy the needed conditions: a pure entangled steady state, with dynamics that does not preserve the even parity subspace.}.
The first
is an effective TMS reservoir realized either via distributed two-mode squeezing \cite{Agusti-arxivquant-ph-2022-t,Andres-Juanes-arxivquant-ph-2025-f} or parametric modulation \cite{Govia-physRevRes-2022-d}, and described by Eq.~\eqref{eq:TMSLindblad}.
In this case, the dark state in Eq.~\eqref{eq:dark-state} has $v = \sinh r / \sqrt{ \cosh(2r)}$, and the Schmidt basis is the computational basis.  
The second example combines collective loss from a directional (chiral) waveguide along with coherent Rabi driving of the communication qubits \cite{Stannigel-newJPhys-2012-i,Irfan-arxivquant-ph-2025-w}:
\begin{equation}
  \mathcal{L}_{\text{Rabi}}\rho = -i[\hat H_{\rm dr} +\hat H_{\rm chi},\rho] + \gamma\mathcal{D}[\hat\sigma_A^- + \hat\sigma_B^-]\rho, \label{eq:Rabi}
\end{equation}
where $\hat H_{\rm dr} = \Omega(\hat\sigma_A^x + \hat\sigma_B^x)$ is the local driving Hamiltonian and $\hat H_{\rm chi} = -\frac{i\gamma}{2}(\hat\sigma_B^+\hat\sigma_A^- - \mathrm{h.c.})$ describes the effective chiral interaction induced by the waveguide. Written in the computational basis, the dark steady state of Eq.~\eqref{eq:dark-state} is
\begin{equation}
  \ket{\psi^c_{ss}} = \frac{1}{\mathcal{N}}\left[\ket{g^c_A g^c_B} + \frac{2\Omega}{i\gamma}\bigl(\ket{e^c_A g^c_B}-\ket{g^c_A e^c_B}\bigr)\right].  \label{eq:Rabi-dark}
\end{equation}
A local $\hat\sigma^x$ rotation on each note by angle $\tan\theta = 4\Omega/\gamma$ brings it to the Schmidt form of \cref{eq:dark-state}, with the entanglement parameter $v$ set by $\Omega/\gamma$ (see \cite{supp}).

For each of these choices of entangling reservoir, the convergence rate $\Gamma$ depends on the choice of time intervals $(\tau_{\rm stb}, \tau_{\rm conv})$. The role of $\tau_{\rm stb}$ is straightforward: it needs to exceed the relaxation time $\tau_{\rm ent}$ of $\mathcal{L}_{\text{ent}}$ (i.e.~inverse dissipative gap), so that the communication qubits are reset to $\ket{\psi^c_{ss}}$ at the start of each subroutine $\mathcal{E}_{ZZ},\mathcal{E}_{XX}$. Numerics confirm that $\Gamma$ saturates once $\tau_{\rm stb}$ is made longer than a few times $\tau_{\rm ent}$. In the relevant operating regimes, $\tau_{\rm stb}\gtrsim 5 \tau_{\rm ent}$ would suffice for both protocols (see \cite{supp}). 

%%%%%%%%%%%%%%%%%%%%%%%%%%
\begin{figure}[t]
    \centering
    \includegraphics[]{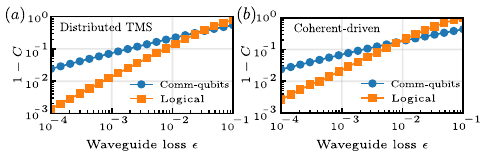}
    \caption{Optimal steady-state concurrence error $1 - \mathcal{C}$ versus waveguide loss $\epsilon$ for communication-qubit-only stabilization via $\mathcal{L}_{\rm ent}$ (blue circles),  and the full logical Floquet protocol (orange squares).  The two panels are for the TMS and Rabi schemes, with $(\tau_{\rm stb}, \tau_{\rm conv})$ optimized at each $\epsilon$. Both Floquet protocols exhibit linear scaling in the small-loss regime whereas communication-qubit-only stabilization scales sub-linearly $\sim \epsilon^{1/2}$.  Below $\epsilon \sim 10^{-2}$ the logical protocol outperforms the $\mathcal{L}_{\rm ent}$-only protocol.}
    \label{fig-3:loss-resillience}
\end{figure}
%%%%%%%%%%%%%%%%%%%%%%%%%%

In contrast, the dependence of $\Gamma$ on the conversion time $\tau_{\rm conv}$ exhibits a non-trivial optimum [Fig.~\ref{fig-2:numerics} (b)]. For the TMS scheme, $\mathcal{L}_{\text{TMS}}$ has weak parity symmetry~\cite{Buca-newJPhys-2012,Albert-physRevA-2014}, and we can derive an analytic expression for 
$p_{\text{conv}}(\tau_{\rm conv})$ by counting quantum jumps (see \cite{supp}).  
Maximizing $p_{\rm conv}(\tau_{\rm conv})$ yields an exact expression for the optimal 
$\tau_{\rm conv}^*$ which agrees with full master-equation simulations (see \cref{fig-2:numerics}(b)). For the Rabi scheme, the absence of weak symmetry leads to a more intricate landscape of $p_{\rm conv}$ \cite{supp}. Regardless, comparable optimal rates are achievable across a broad range of $\Omega/\gamma$ . We find that $\Gamma_{\max}$ increases monotonically with the bath entanglement parameter $v$ (see \cref{fig-2:numerics}(a)), as expected for distillation dynamics.  

{\it Robustness to communication loss.}--- 
Implementations of our remote protocol will primarily be limited by photon loss in the waveguide used to realize the entangling dissipation.  We parameterize this by a power loss $\epsilon\equiv 1-|\eta|^2$, where $\eta$ is the transmission amplitude through the waveguide.  Loss corrupts the entangling bath, modifying the Lindbladian it generates to $\mathcal{L}^{\epsilon}_{\text{ent}}$, and causing its steady state to become impure~\cite{supp}.  As discussed in the End Matter, this yields two distinct error mechanisms, both of which scale like $\epsilon$ to leading order. 

We quantify the stabilized logical entanglement in the presence of waveguide loss using a strict entanglement monotone, the steady-state logical concurrence $\mathcal{C}$; 
based on the above discussion, we expect a linear concurrence error $1-\mathcal{C}\propto\epsilon$ at small loss. \cref{fig-3:loss-resillience} confirms this linear scaling across nearly two decades of $\epsilon$ for both the TMS and Rabi schemes, with $(\tau_{\rm stb},\tau_{\rm conv})$ optimized at each $\epsilon$. Strikingly, in the small-loss regime our Floquet reservoir engineering protocol achieves a concurrence error substantially below the optimal bare physical-qubit concurrence (i.e.~the best entanglement achievable directly from $\mathcal{L}^\epsilon_{\rm ent}$ after optimizing over the entangling parameters $r$ or $\Omega/\gamma$).  While this optimized communication-qubit concurrence error scales like $\propto \epsilon^{1/2}$ (something that can be related to the time-entanglement tradeoff of Ref.~\cite{Pocklington-prxQuantum-2024-w}), our Floquet scheme achieves the much-improved scaling $\propto \epsilon$.  
This directly shows that our protocol performs autonomous entanglement {\it distillation}~\cite{Vollbrecht-physRevLett-2011-y}, concentrating the weak, noisy physical entanglement of $\hat\rho^\epsilon_{\rm ss}$ into a near-maximally-entangled logical Bell pair. This
distillation behaviour is most pronounced at small $\epsilon$, degrading for larger $\epsilon$ due to higher-order contributions (with the curves in \cref{fig-3:loss-resillience} crossing at a finite value of $\epsilon$). 

%%%%%%%%%%%%%%%%%%%%%%%%%%%%%%%%
\begin{figure}[tb]
  \centering
  \includegraphics{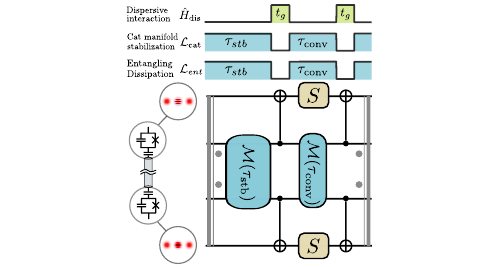}
  \caption{\textbf{Stabilization of remote cat-qubit Bell state.}
    Pulse sequence (top) and circuit diagram (bottom) for one stabilization cycle, using the Rabi-driven entangling reservoir $\mathcal{L}_{\rm ent}$ [\cref{eq:Rabi}]. Within a cycle, $\mathcal{L}_{\rm ent}$ is activated on the transmons for two time intervals $(\tau_{\rm stb}, \tau_{\rm conv})$, interleaved with dispersive CNOT gates of duration $t_g$ and single-photon-driven cat phase gates $S$; the two-photon cat stabilization $\mathcal{L}_{\rm cat}$ acts continuously to protect the logical states.}
  \label{fig-4:cat-bell-stabilization}
\end{figure}
%%%%%%%%%%%%%%%%%%%%%%%%%%%%%%%%

{\it Experimental realization.}--- Inspired by recent experiments~\cite{Putterman-nature-2025-x,Reglade-nature-2024-l,Leghtas-science-2015,Ofek-nature-2016-y}, we now map our protocol onto a hybrid continuous-variable--discrete-variable (CV-DV) superconducting architecture
\cite{Liu-prxQuantum-2026-d}, using bosonic cat qubits as the protected logical memories and transmons as the physical communication qubits. The logical codewords are orthogonalized coherent states $\ket{0^L}\approx\ket{\alpha}$ and $\ket{1^L}\approx\ket{-\alpha}$ 
\footnote{The exactly orthogonal states spanning the steady-state
manifold of $\mathcal{L}_{\rm cat}$ are the even- and odd-parity cat
states $|\mathcal{C}_\alpha^\pm\rangle = (|\alpha\rangle \pm
|{-\alpha}\rangle)/\mathcal{N}_\pm$. The code words are their equal
superpositions, $|{0^L}\rangle = (|\mathcal{C}_\alpha^+\rangle +
|\mathcal{C}_\alpha^-\rangle)/\sqrt{2}$ and $|{1^L}\rangle =
(|\mathcal{C}_\alpha^+\rangle - |\mathcal{C}_\alpha^-\rangle)/\sqrt{2}$,
which coincide with the coherent states $|{\pm\alpha}\rangle$ up to
corrections of order $e^{-2|\alpha|^2}$.}, 
stabilized within the cat manifold by two-photon driven dissipation $\mathcal{L}_{\rm cat} = \kappa_2 \mathcal{D}[\hat a^2 - \alpha^2]$ \cite{Leghtas-science-2015}, where $\hat{a}$ is the photon annihilation operator for the storage cavity. The remote entangling reservoir $\mathcal{L}_{\rm ent}$ acting on the two transmons is taken to be the Rabi-driven scheme of \cref{eq:Rabi}.

The cat encoding admits only operations that preserve its exponential $Z$-noise bias. A native bias-preserving two-qubit gate on this platform is a transmon-controlled-$X$ on the cat, implemented in a recent QEC experiment~\cite{Putterman-nature-2025-x} by a $\chi$-matched dispersive coupling $\hat H_{\rm dis} = \hat a^\dagger \hat a \otimes (\chi_{ge}\ket{e}\bra{e} + \chi_{gf}\ket{f}\bra{f})$ with $\chi_{ge} = \chi_{gf} \equiv \chi$, over a gate time $t_g = \pi/\chi$. To deal with finite gate time, we assume the entangling dissipation can be turned off by using a tunable coupler or via destructive interference~\cite{Almanakly-arxivquant-ph-2026-d}. To achieve a complete set of noise biased operations, we replace the $\{CX, CZ\}$ gate set
in \cref{fig-1:Schematic} with $\{CX, CY\}$, realizing $CY$ from $CX$ dressed by a single-photon-driven cat phase gate $\hat S$ \cite{Guillaud-physRevX-2019-a}. 

At first glance,~\cref{fig-4:cat-bell-stabilization}(b) 
does not seem equivalent to our proposed protocol in~\cref{fig-1:Schematic} and the channel in Eq.~\eqref{eq:composite-channel}.  Remarkably, it is equivalent when repeated twice, i.e.~$\mathcal{E}_{\rm cat}^2 = \mathcal{E}_{YY} \circ (\hat Z^L_A \otimes \hat Z^L_B\otimes \mathbb{I}_c)\circ \mathcal{E}_{XX}$ (where the extra $\hat{Z}^L$ play no role, see~\cite{supp}).  Apart from direct relevance to the dissipative cat platform, this also represents another method for implementing our general scheme.
The protocol then stabilizes the logical Bell state fixed by $\langle X^L_A X^L_B\rangle = \langle Y^L_A Y^L_B\rangle = 1$ 
which is local-unitary equivalent to the $\ket{\Phi^L_+}$ target state discussed above. More details on this implementation are presented in the End Matter.

{\it Conclusion.}---  We have introduced a general Floquet reservoir engineering framework for deterministically stabilizing a maximally-entangled Bell state between two remote protected qubits, using nothing more than a communication-qubit entangling reservoir together with local controlled-logical gates.  The protocol autonomously mimics the physics of a measurement-plus-feedback protocol, without any need for explicit measurements or classical communication.   
It also acts as an autonomous distillation engine: it concentrates the weak, noisy entanglement that the bath sustains between the communication qubits into a highly entangled logical Bell pair that the bath itself never touches.  Because the construction treats the logical encoding as a black box, it extends naturally from cat qubits to GKP and other codes, charting a route toward distributed, fault-tolerant quantum networks assembled from entangling reservoirs never designed to act on logical information.

{\it Acknowledgments.}--- We thank Liang Jiang and Pei Zeng for insightful discussions. This work was supported by the Army Research Office under Grant No. W911NF-23-1-0077, and by the Simons Foundation through a Simons Investigator Award (Grant No. 669487). 

\bibliography{ref}

\newpage
\appendix*
\section{End Matter}

\begin{figure}[h!]
    \centering
    \includegraphics[]{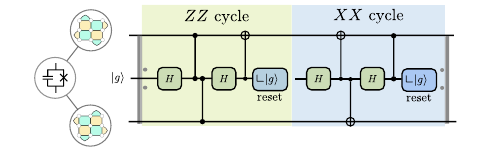}
    \caption{\textbf{Reference local parity measurement and feedback protocol.} Two co-located logical qubits (top and bottom rails) share a single physical ancilla initialized in $\ket{g}$ (middle rail). In the $ZZ$ cycle, two Hadamard gates sandwich a pair of controlled-$Z^L$ gates, mapping the joint logical parity onto the ancilla population; an ancilla-controlled $X^L$ corrects the parity, and the ancilla is then reset. The $XX$ cycle acts analogously in the rotated basis.}
    \label{end-fig-1:local-measurement}
\end{figure}

{\it Comparison to explicit measurement and feedback dynamics.}---
For two co-located logical qubits, stabilization via parity measurement and feedback is well established~\cite{Riste-nature-2013, Roch-physRevLett-2014, Andersen-npjQuantumInf-2019-a}: a single shared ancilla coherently records the joint logical parity in its population, controls the corrective flip, and is then reset, as depicted in \cref{end-fig-1:local-measurement}. This reference protocol is deterministic, correcting each parity error in a single shot, i.e.~$p_{\text{conv}} = 1$ in~\cref{eq:zz-recursion}.
%%%%%%%%%%%%%%%%%%%%%%
%%%%%%%%%%%%%%%%%%%%%%

Moving to the remote setting raises two fundamental obstacles, both of which our protocol overcomes. First, the remote parity measurement needed for each step is no longer free, as it consumes 1 ebit of entanglement per shot \cite{Eisert-physRevA-2000, Bandyopadhyay-physRevA-2009}, more than the weakly entangled bath resource $\ket{\psi^c_{ss}}$ carries. We circumvent this in our protocol by tolerating an imperfect conversion: the non-local dissipation itself transduces the phase into locally readable parity with finite probability $p_{\text{conv}} < 1$. Second, remote measurement and feedback would require classical communication of the outcome, introducing a closed-loop delay that precludes autonomous operation. We avoid this through a subtler route: rather than measuring the remote parity directly, we let the non-local dissipation itself verify the parity, converting global parity information into a local signal on the communication qubits.

\begin{figure}[h!]
    \centering
    \includegraphics{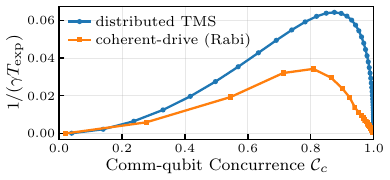}
    \caption{\textbf{Typical protocol run time $T_{\rm exp}$ as a function of the communication-qubit entanglement.} The non-monotonic dependence yields a clear optimum, set by the tradeoff between larger conversion rate $p_{\rm conv}$ and slower stabilization $\tau_{\rm ent}$ as the entanglement increases.}
    \label{end-fig-2:total-time}
\end{figure}

{\it Typical time scale.}---Another experimentally relevant quantity is the typical run time $T_{\rm exp}$ of the protocol. Taking all local gates to be instantaneous, it is set by the time per cycle and the number of cycles $N_{\rm exp}$ needed to reach the steady state,
\begin{equation}
    T_{\rm exp} = N_{\rm exp} \times 2(\tau_{\rm stb} + \tau_{\rm conv}) = \frac{2(\tau_{\rm stb} + \tau_{\rm conv})}{\Gamma},
\end{equation}
where $\Gamma$ is the stabilization rate defined in the main text and the factor of two accounts for the two subroutines per cycle. As shown in \cref{end-fig-2:total-time}, $T_{\rm exp}$ depends non-monotonically on the communication-qubit entanglement and exhibits a clear optimum. This optimum reflects the time-entanglement tradeoff of the entangling reservoir: stronger bath entanglement boosts the conversion but slows the stabilization, so an intermediate value minimizes the total run time.

{\it Loss degradation mechanisms.}---
Loss corrupts the entangling bath, modifying the Lindbladian to $\mathcal{L}^{\epsilon}_{\text{ent}}$, and causing its steady state $\hat \rho^\epsilon_{\rm ss}$ to become impure and have a small odd-parity component (probability $O(\epsilon)$) (see \cite{supp}).  
Waveguide loss has two deleterious effects, both linear in $\epsilon$. First, during the 
$\mathcal{E}_{ZZ}$ ($\mathcal{E}_{XX}$) subroutine the $XX$ ($ZZ$) logical correlator decays.  
This is a direct consequence of the unwanted odd-parity component in $\hat \rho^\epsilon_{\rm ss}$.   
For example, the $\mathcal{E}_{XX}$ no longer leaves $ZZ$ unchanged, but causes additional parity flip during step (ii) of our protocol, where the detrimental odd-parity component causes unwanted single $X$ operation through $\mathcal{C}_X$. This effect reduces cleanly in the $\tau_{\rm stb} \gg \tau_{\rm ent}$ limit, namely
\begin{equation}
  \langle \mathcal{E}^\dagger_{XX}( \hat Z^L_A \hat Z^L_B )\rangle = \langle \hat \sigma^z_A \hat \sigma^z_B\rangle_{\rm ss}\, \langle \hat Z^L_A \hat Z^L_B\rangle,
\end{equation}
where $\langle \hat \sigma^z_A \hat \sigma^z_B\rangle_{\rm ss} = \tr(\hat \sigma^z_A \hat \sigma^z_B \hat \rho^\epsilon_{\rm ss}) = 1-O(\epsilon)$.

Second, the $O(\epsilon)$ odd-parity component in $\hat \rho^\epsilon_{\rm ss}$  produces {\it false positives} and unwanted feedback during step (iv) of our protocol: 
even if the logical parity is even, the corrupted communication qubit state can trigger an unwanted parity flip of the logical qubits.  
This happens at a rate $p_{\text{err}} \equiv p(P^c_{\rm odd} \mid P^L_{\rm even})$, with $P^c_{\rm odd}$ and $P^L_{\rm even}$ the communication- and logical-qubit odd(even) parity subspaces. This false-positive rate competes with the true conversion probability $p_{\text{conv}}$ and modifies the operator dynamics under $\mathcal{E}_{ZZ}$ to
\begin{equation}
\begin{split}
  \langle \mathcal{E}_{ZZ}^\dagger (\hat Z_A^L \hat Z_B^L)\rangle = {} & p_{\text{conv}}-p_{\text{err}} \\
  & + (1-p_{\text{conv}}-p_{\text{err}})\, \langle \hat Z_A^L \hat Z_B^L \rangle.
\end{split}
\end{equation}
This error will also contribute to the steady state as $O(\epsilon)$.  

Our distillation advantage originates from the less favorable error scaling of
schemes that use only communication qubits. Their sub-linear scaling
$\Delta F^* \sim \epsilon^{1/2}$ is a fundamental limit whenever the
time-entanglement tradeoff applies.  We provide a heuristic argument for this in what follows.

When the designed steady state is close to a maximally entangled state,
$\Delta F_{\rm ss} = 1 - F_{\rm ss} \ll 1$, the steady-state infidelity and
the rate of reaching the steady state satisfy~\cite{Pocklington-prxQuantum-2024-w}
\begin{equation}
    \Delta F_{\rm ss} = c\, \lambda_{\rm ent},
\end{equation}
where $\lambda_{\rm ent} = 1/(\gamma \tau_{\rm ent})$ is the dimensionless gap of
$\mathcal{L}_{\rm ent}$ and $c$ is a constant. Now consider noise of strength $\epsilon$. In the regime
$\lambda_{\rm ent} \gg \epsilon$, first-order perturbation theory applies and
the noisy steady state reads
$\rho_\epsilon = \rho_{\rm ss} + \frac{a\epsilon}{\lambda_{\rm ent}}\rho_1$,
where $\rho_1$ is traceless and $a$ is a constant set by the noise Lindbladian.
The infidelity of the noisy steady state therefore has two parts,
\begin{equation}
    \Delta F(\epsilon) = c\, \lambda_{\rm ent}
    + \frac{a\, \Delta F_1\, \epsilon}{\lambda_{\rm ent}},
\end{equation}
where $\Delta F_1$ is the infidelity contribution from $\rho_1$. The first term
is the error of the ideal target steady state itself, and the second is the error induced by
the noise. Since $c$, $a$, and $\Delta F_1$ are all independent of $\epsilon$,
the two terms compete: increasing the gap suppresses the noise-induced error
but raises the designed-state error. Minimizing over $\lambda_{\rm ent}$, which
in practice means tuning the entangling parameter, gives the optimal gap
$\lambda_{\rm ent}^* \sim \epsilon^{1/2}$ and hence the optimal infidelity
$\Delta F^*(\epsilon) \sim \epsilon^{1/2}$. We note that when $\Delta F \ll 1$, the concurrence deficit $1 - C$ scales in the same way as the infidelity $1 - F$, so the $\epsilon^{1/2}$ scaling holds for either entanglement measure.

%%%%%%%%%%%%%%%%%%%%%%%%%
\begin{figure}[h!]
    \centering
    \includegraphics[]{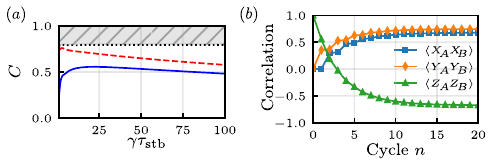}
    \caption{\textbf{Experimental realization of cat Bell state.} (a) Optimized steady-state concurrence $\mathcal{C}$ versus stabilization time $\gamma\tau_{\rm stb}$ at waveguide loss $\epsilon=10^{-2}$. The dotted (black), dashed (red), and solid (blue) curves correspond to transmon $T_1$ loss only, additional cavity dephasing, and the full noise model respectively. The shaded area is inaccessible regime set by the waveguide loss.
  (b) Convergence of the logical correlators over stabilization cycles $n$ under the full noise model with $\gamma \tau_{\rm stb} = 20$, identifying the stabilized Bell state; the comparable decay of $\langle X^L_A X^L_B\rangle$ and $\langle Y^L_A Y^L_B\rangle$ confirms preservation of the noise bias.}
    \label{end-fig-3:cat}
\end{figure}
%%%%%%%%%%%%%%%%%%%%%%%%%

{\it Experimental realization.}--- We assess the experimental feasibility of the cat qubit realization of our scheme, through a full master-equation simulation including waveguide loss, transmon relaxation, cat idling dephasing (cavity photon loss), and finite-gate-time errors (parameters listed in the Supplemental Material). With strong waveguide coupling $\gamma/2\pi = 10\,\text{MHz}$ \cite{Shah-prxQuantum-2024-m} and $\epsilon = 10^{-2}$, waveguide loss dominates over transmon relaxation ($T_{1,t} = 50\,\mu\text{s}$) and sets the entanglement ceiling for the logical Bell state [shaded region in \cref{end-fig-3:cat}(a)]. The residual error reflects 
competing tendencies that determine an optimal stabilization time $\tau_{\rm stb}$.  On the one hand, cat idling dephasing (i.e.~a logical phase-flip rate $\sim |\alpha|^2/T_{1,c}$ set by single-photon loss $T_{1,c} = 200\,\mu\text{s}$ and $|\alpha|^2 = 3$) penalizes long cycles.  On the other hand, the the finite gate time $t_g = \pi/\chi = 500\,\text{ns}$ ($\chi/2\pi = 1\,\text{MHz}$, to be compared with $1/\gamma \approx 16\,\text{ns}$) penalizes the use of short stabilization times $\tau_{\rm stb}$. 
Balancing these two tendencies yields the optimal $\tau_{\rm stb}$ in \cref{end-fig-3:cat}(a). Finally, the $Z$-biased cat noise decays $\langle X^L_A X^L_B\rangle$ and $\langle Y^L_A Y^L_B\rangle$ comparably while $\langle Z^L_A Z^L_B\rangle \to -1$, confirming that the noise bias is preserved across the full stabilization cycle [\cref{end-fig-3:cat}(b)].

%%%%%%%%%%%%%%%%%%% SUPPLEMENT %%%%%%%%%%%%%%%%%%%
% \clearpage
% \pagebreak
% \renewcommand{\theequation}{S\arabic{equation}}
% \renewcommand{\thefigure}{S\arabic{figure}}
% \renewcommand{\thetable}{S\arabic{table}}
% \renewcommand{\thesection}{S\Roman{section}}

\end{document}

\title{Supplemental Material for:
Floquet Reservoir Engineering for Remote Logical Entanglement}

\author{Mingxing Yao}
\email{mxyao@uchicago.edu}
\affiliation{Pritzker School of Molecular Engineering, University of Chicago, Chicago, IL 60637, USA}

\author{Aashish A. Clerk}
\affiliation{Pritzker School of Molecular Engineering, University of Chicago, Chicago, IL 60637, USA}

\maketitle
\tableofcontents
% -------------------------------------------------------
\section{Stabilization Cycle Dynamics}
\label{sec:cycle-dynamics}
% -------------------------------------------------------

The protocol stabilizes the two-logical-qubit Bell state $\ket{\Phi^L_+}=(\ket{0^L_A 0^L_B}+\ket{1^L_A 1^L_B})/\sqrt{2}$ by alternating between a $Z^L Z^L$-correction cycle that pumps $\langle \hat Z^L_A\hat Z^L_B\rangle\to 1$ and an $X^L X^L$-correction cycle that pumps $\langle\hat X^L_A\hat X^L_B\rangle\to 1$. These two stabilizers commute and together uniquely fix $\ket{\Phi^L_+}$, so the analysis reduces to tracking each correlator through one cycle. In this section we walk through the joint state evolution to expose the role of the entangling bath as a global-to-local information converter, derive the rate equations quoted in the main text, and pin the convergence rate to the conversion probability $p_{\rm conv}$. Throughout we work in the Schmidt basis $\{\ket{0^c_A},\ket{1^c_A}\}$ of the communication qubit at node A (analogously for node B), in which the dark state of the entangling Lindbladian takes the rank-one form $\ket{\psi^c_{\rm ss}} = u\ket{0^c_A 0^c_B}+v\ket{1^c_A 1^c_B}$ with $u,v\ge 0$ and $u^2+v^2=1$. Pauli operators $\hat\sigma^\alpha_{A(B)}$ act on the communication qubit at node A (B); logical operators are always labeled with the superscript $L$.

\subsection{Cycle structure}
\label{sec:cycle-stages}

A $Z^L Z^L$-correction cycle is the composition of four channels acting on the joint logical-plus-communication system,
\begin{equation}
\mathcal{E}_{ZZ} \;=\; \mathcal{C}_X\,\circ\,\bigl[\mathbb{I}_L\otimes\mathcal{M}(\tau_{\rm conv})\bigr]\,\circ\,\mathcal{C}_Z\,\circ\,\bigl[\mathbb{I}_L\otimes\mathcal{M}(\tau_{\rm stb})\bigr],
\label{eq:cycle-channel}
\end{equation}
applied to the input of the joint communication and logical state. The two unitary channels
\begin{equation}
\mathcal{C}_Z[\,\cdot\,] = U_{CZ}(\cdot)U_{CZ}^\dagger,\quad U_{CZ} = CZ_{A\to A}\,CZ_{B\to B},\qquad \mathcal{C}_X[\,\cdot\,] = U_{CX}(\cdot)U_{CX}^\dagger,\quad U_{CX} = CX_{A\to A}\,CX_{B\to B},
\end{equation}
are products of two communication-qubit-controlled logical gates, one at each node (the second index labels the logical target). The $X^L X^L$-correction cycle $\mathcal{E}_{XX}$ has the same structure with $\mathcal{C}_Z\leftrightarrow\mathcal{C}_X$.

We work in the limit $\tau_{\rm stb}\to\infty$ throughout this section, so that the resource-preparation stage drives the communication qubits to the pure dark state, $\mathcal{M}(\tau_{\rm stb})(\hat\rho^c_0)\to\ket{\psi^c_{\rm ss}}\bra{\psi^c_{\rm ss}}$. Finite-$\tau_{\rm stb}$ corrections, which set the minimum cycle time, are discussed in \cref{sec:saturation}.

\subsection{State-level walkthrough}
\label{sec:walkthrough}

The only step in $\mathcal{E}_{ZZ}$ that requires the non-local resource is the bath verification $\mathcal{M}(\tau_{\rm conv})$. To see precisely what this step does, we walk through a $Z^L Z^L$ cycle with a general pure logical input $\ket{\psi^L} = a\ket{00}+b\ket{01}+c\ket{10}+d\ket{11}$ (the same statements hold at the density-matrix level by linearity).

After resource preparation the joint state is
\begin{equation}
\ket{\Psi_1} = \ket{\psi^L}\otimes\ket{\psi^c_{\rm ss}}.
\end{equation}
The parity-encoding step $\mathcal{C}_Z$ attaches a $-$ sign to every component in which a logical qubit is in $\ket{1^L}$ and the corresponding communication qubit is in $\ket{1^c}$. Grouping the logical state by its $Z$-parity, the joint state becomes
\begin{equation}
\ket{\Psi_2} = \underbrace{\bigl(a\ket{00}+d\ket{11}\bigr)}_{Z^L Z^L = +1}\otimes\ket{\psi^c_{\rm ss}}\;+\;\underbrace{\bigl(b\ket{01}+c\ket{10}\bigr)}_{Z^L Z^L = -1}\otimes\ket{\psi^c_{\rm det}},
\label{eq:Psi2}
\end{equation}
where we have defined the orthogonal partner state
\begin{equation}
\ket{\psi^c_{\rm det}} \equiv u\ket{0^c_A 0^c_B} - v\ket{1^c_A 1^c_B}.
\end{equation}
The even-parity logical branch remains the dark state, while the odd-parity branch becomes $\ket{\psi^c_{\rm det}}$. The two communication states differ only by a global sign on the $\ket{1^c 1^c}$ component, and crucially they have identical reduced density matrices on each individual communication qubit: $\rho^c_A = \rho^c_B = u^2\ket{0^c}\bra{0^c}+v^2\ket{1^c}\bra{1^c}$. No local measurement on $A$ or $B$ alone can distinguish $\ket{\psi^c_{\rm ss}}$ from $\ket{\psi^c_{\rm det}}$, which is why no local feedback decision is possible at this stage --- the parity information lives in a non-local relative phase between the two nodes.

The bath verification step $\mathcal{M}(\tau_{\rm conv})$ converts that non-local phase into a local parity bit. Because $\ket{\psi^c_{\rm ss}}$ is the unique dark state of $\mathcal{L}_{\rm ent}$, the even-parity branch is stationary under $\mathcal{M}(\tau_{\rm conv})$. The error state $\ket{\psi^c_{\rm det}}$, on the other hand, has nonzero overlap with the bright modes of $\mathcal{L}_{\rm ent}$ and relaxes toward $\ket{\psi^c_{\rm ss}}$ via quantum jumps. Each jump of the entangling dissipator changes the joint communication parity by one, so with probability
\begin{equation}
p_{\rm conv} \equiv \tr\!\left[\hat P^c_{\rm odd}\,\mathcal{M}(\tau_{\rm conv})\bigl(\ket{\psi^c_{\rm det}}\bra{\psi^c_{\rm det}}\bigr)\right],\qquad \hat P^c_{\rm odd} = \ket{0^c_A 1^c_B}\bra{0^c_A 1^c_B} + \ket{1^c_A 0^c_B}\bra{1^c_A 0^c_B},
\label{eq:pdet-def}
\end{equation}
the communication register ends the verification step in the odd-parity subspace. The conversion probability $p_{\rm conv}$ is the central figure of merit of the protocol; we compute it analytically for the TMS dissipator in \cref{sec:tms-pdet}. Operationally, $\mathcal{M}(\tau_{\rm conv})$ leaves the dark-state branch alone and partially relaxes the error-state branch into a state whose local communication parity is now visible: a logical $Z^L Z^L$ error has been heralded by a flipped transmon parity, with success rate $p_{\rm conv}$ per cycle.

The final step $\mathcal{C}_X$ uses this heralded parity as a coherent control. When the communication qubits sit in the odd-parity sector at the start of $\mathcal{C}_X$, $U_{CX}$ flips both logical qubits and maps $b\ket{01}+c\ket{10}\to b\ket{11}+c\ket{00}$, returning the logical state to the even-parity sector. When the verification fails to detect (communication qubits stay in even parity), $\mathcal{C}_X$ acts as $X^L_A X^L_B$ on the logical state, which commutes with $\hat Z^L_A\hat Z^L_B$ and therefore does no damage. The cycle is self-correcting on detection and idle on non-detection --- no classical measurement and feedback is required, and the only resource consumed each cycle is the non-local bath activation $\mathcal{M}(\tau_{\rm stb})$ and $\mathcal{M}(\tau_{\rm conv})$.

\subsection{Equation of motion for \texorpdfstring{$\langle\hat Z^L_A\hat Z^L_B\rangle$}{<ZZ>}}
\label{sec:ZZ-eom}

The walkthrough already encodes the rate equation. Let $p_e^{(n)}$ and $p_o^{(n)} = 1-p_e^{(n)}$ denote the populations of the even and odd logical-parity sectors in $\hat\rho_n$, so that $\langle\hat Z^L_A\hat Z^L_B\rangle_n = p_e^{(n)}-p_o^{(n)}$. From \cref{eq:Psi2}, $\mathcal{C}_Z$ preserves the logical parity blocks and entangles the even (odd) block exclusively with $\ket{\psi^c_{\rm ss}}$ ($\ket{\psi^c_{\rm det}}$). The subsequent $\mathcal{M}(\tau_{\rm conv})$ keeps the dark-state branch in the even communication-parity subspace with certainty, and pushes the error-state branch into the odd communication-parity subspace with probability $p_{\rm conv}$. The feedback $\mathcal{C}_X$ then flips the logical parity if and only if the communication qubits show odd parity at the moment of the gate. Reading off the transition probabilities,
\begin{align}
p_e^{(n+1)} &= 1\cdot p_e^{(n)} \;+\; p_{\rm conv}\cdot p_o^{(n)},\\
p_o^{(n+1)} &= 0\cdot p_e^{(n)} \;+\; (1-p_{\rm conv})\cdot p_o^{(n)},
\end{align}
where the four coefficients are the conditional probabilities of the joint event (no logical flip | logical even), (logical flip | logical odd), (logical flip | logical even), (no logical flip | logical odd), respectively. Subtracting these two equations and using $p_o^{(n)} = (1-\langle\hat Z^L_A\hat Z^L_B\rangle_n)/2$ yields the central result
\begin{equation}
{\;\langle\hat Z^L_A\hat Z^L_B\rangle_{n+1} \;=\; \langle\hat Z^L_A\hat Z^L_B\rangle_n \;+\; p_{\rm conv}\bigl(1-\langle\hat Z^L_A\hat Z^L_B\rangle_n\bigr).\;}
\label{eq:ZZ-result}
\end{equation}
The fixed point is $\langle\hat Z^L_A\hat Z^L_B\rangle_\infty = 1$, reached exponentially:
\begin{equation}
1-\langle\hat Z^L_A\hat Z^L_B\rangle_n = \bigl(1-\langle\hat Z^L_A\hat Z^L_B\rangle_0\bigr)e^{-\Gamma n},\qquad \Gamma \equiv -\ln(1-p_{\rm conv}) \approx p_{\rm conv}\;\text{ for }p_{\rm conv}\ll 1.
\label{eq:ZZ-exponential}
\end{equation}
Two qualitative features of \cref{eq:ZZ-result} are worth emphasizing. First, perfect detection $p_{\rm conv}=1$ is \emph{not} required for the steady-state correlator to reach unity --- any $p_{\rm conv}>0$ pumps $\langle\hat Z^L_A\hat Z^L_B\rangle$ all the way to $+1$. This is the qualitative point that distinguishes the present scheme from direct dissipative entanglement of physical qubits, where the steady-state physical correlator saturates at a value strictly below $1$ for any finite resource parameter. Second, the only role of $p_{\rm conv}$ in the ideal protocol is to set the convergence \emph{rate} $\Gamma$; tuning the bath-cycle parameters $\tau_{\rm stb}$, $\tau_{\rm conv}$, and the resource strength of $\mathcal{L}_{\rm ent}$ to maximize $\Gamma$ is therefore the central optimization task.

\subsection{Equation of motion for \texorpdfstring{$\langle\hat X^L_A\hat X^L_B\rangle$}{<XX>}}
\label{sec:XX-eom}

Independent stabilization of the two correlators requires that the $Z^L Z^L$ cycle leave $\langle\hat X^L_A\hat X^L_B\rangle$ undisturbed (and vice versa). The cleanest route to this statement is the Heisenberg picture for $\mathcal{E}_{ZZ}$. Using
\begin{equation}
U_{CZ}^\dagger\,\hat X^L_A\hat X^L_B\,U_{CZ} = \hat\sigma^z_A\hat\sigma^z_B\,\hat X^L_A\hat X^L_B,\qquad U_{CX}^\dagger\,\hat X^L_A\hat X^L_B\,U_{CX} = \hat X^L_A\hat X^L_B.
\end{equation}
The Heisenberg-evolved observable factorizes between the logical and communication sectors, and
\begin{align}
\langle\hat X^L_A\hat X^L_B\rangle_{n+1}
&= \tr\!\left[\hat X^L_A\hat X^L_B\,\mathcal{E}_{ZZ}(\hat\rho_n\otimes\hat\rho^c_{\rm ss})\right]\nonumber\\
&= \tr_L\!\bigl[\hat X^L_A\hat X^L_B\,\hat\rho_n\bigr]\cdot\tr_c\!\left[\hat\sigma^z_A\hat\sigma^z_B\,\mathcal{M}(\tau_{\rm conv})(\hat\rho^c_{\rm ss})\right]\nonumber\\
&= \langle\hat X^L_A\hat X^L_B\rangle_n\cdot\langle\hat\sigma^z_A\hat\sigma^z_B\rangle_{\rm ss}.
\label{eq:XX-during-ZZ}
\end{align}
In the last line we used $\mathcal{M}(\tau_{\rm conv})(\hat\rho^c_{\rm ss}) = \hat\rho^c_{\rm ss}$. In the ideal case $\hat\rho^c_{\rm ss} = \ket{\psi^c_{\rm ss}}\bra{\psi^c_{\rm ss}}$ and the Schmidt-basis dark state has both qubits perfectly correlated, so $\langle\hat\sigma^z_A\hat\sigma^z_B\rangle_{\rm ss} = 1$ and $\langle\hat X^L_A\hat X^L_B\rangle_{n+1} = \langle\hat X^L_A\hat X^L_B\rangle_n$ exactly: the $Z^L Z^L$ cycle is transparent to the $X^L X^L$ correlator. By the symmetric calculation, $\mathcal{E}_{XX}$ leaves $\langle\hat Z^L_A\hat Z^L_B\rangle$ invariant and pumps $\langle\hat X^L_A\hat X^L_B\rangle\to 1$ at the same exponential rate $\Gamma$. A full protocol cycle, $\mathcal{E}_{XX}\circ\mathcal{E}_{ZZ}$, therefore drives both correlators to unity independently, with the joint convergence rate set by the single quantity $p_{\rm conv}$.

The Bell-state lower bound $F\ge(\langle\hat X^L_A\hat X^L_B\rangle+\langle\hat Z^L_A\hat Z^L_B\rangle)/2$ inherits the same exponential approach, and the logical infidelity decays as $1-F\propto e^{-\Gamma n}$, completing the derivation of the central rate equations quoted in the main text. We will later turn in \cref{sec:tms-pdet} to evaluating $p_{\rm conv}$ analytically for the TMS dissipator. Next we introduce the two main dissipative channels we considered to implement the resource $\mathcal{M}(\tau)$ in \cref{sec:M-realization}.

% -------------------------------------------------------
\section{Experimental realization of \texorpdfstring{$\mathcal{M}(\tau)$}{M(tau)}}
\label{sec:M-realization}
% -------------------------------------------------------

In this section, we make the entangling Lindbladian $\mathcal{L}_{\rm ent}$ mentioned in \cref{sec:cycle-dynamics} concrete by recalling the two paradigmatic remote dissipative entanglement schemes used in the main text --- the two-mode squeezing (TMS) dissipator and the Rabi-driven cascade dissipator. We derive their modifications under finite transmission loss in the communication channel.

\subsection{Two-mode squeezing dissipator}
\label{sec:impl-tms}

The TMS scheme drives the two communication qubits with a shared two-mode squeezed reservoir. In the parametrically engineered limit \cite{Agusti-arxivquant-ph-2022-t, Andres-Juanes-arxivquant-ph-2025-f}, the joint master equation takes the form
\begin{equation}
\mathcal{L}_{\rm TMS}\hat\rho = \gamma\,\mathcal{D}\!\bigl[\cosh r\,\hat\sigma^-_A - \sinh r\,\hat\sigma^+_B\bigr]\hat\rho + \gamma\,\mathcal{D}\!\bigl[\cosh r\,\hat\sigma^-_B - \sinh r\,\hat\sigma^+_A\bigr]\hat\rho,
\label{eq:TMS-clean}
\end{equation}
where $r$ is the squeezing parameter set by the parametric pump strength and $\gamma$ is the overall dissipative rate. The two jump operators are Bogoliubov-transformed lowering operators of the joint system; their unique simultaneous dark state is the two-qubit two-mode squeezed vacuum, written in the computational basis $\{\ket{g^c},\ket{e^c}\}$ as
\begin{equation}
\ket{\psi^c_{\rm ss}} = u\ket{g^c_A g^c_B} + v\ket{e^c_A e^c_B},\qquad u = \frac{\cosh r}{\sqrt{\cosh 2r}},\quad v = \frac{\sinh r}{\sqrt{\cosh 2r}}.
\label{eq:TMS-dark}
\end{equation}
For the TMS dissipator the Schmidt basis of $\ket{\psi^c_{\rm ss}}$ coincides with the computational basis at each node, so the Schmidt-basis identification $\ket{0^c}\equiv\ket{g^c}$, $\ket{1^c}\equiv\ket{e^c}$ used in \cref{sec:cycle-dynamics} requires no local rotation and the cycle of \cref{sec:cycle-dynamics} is implemented directly in the bare communication-qubit basis. The bare-qubit steady-state concurrence increases monotonically with $r$ while saturating strictly below $1$ at any finite $r$, meanwhile the dissipative gap closes as the Bell state fidelity approaches $1$ \cite{Pocklington-prxQuantum-2024-w}.

\subsection{Rabi-driven cascade dissipator}
\label{sec:impl-rabi}

The Rabi-driven scheme implements remote entanglement via chiral waveguide coupling \cite{Stannigel-newJPhys-2012-i, Irfan-arxivquant-ph-2025-w}: each communication qubit is locally Rabi-driven at amplitude $\Omega$ and decays into a unidirectional waveguide that propagates from node $A$ to node $B$. After eliminating the waveguide mode, the resulting cascaded master equation is
\begin{equation}
\mathcal{L}_{\rm Rabi}\hat\rho = -i\!\left[\Omega(\hat\sigma^x_A+\hat\sigma^x_B) - \tfrac{i\gamma}{2}\bigl(\hat\sigma^+_B\hat\sigma^-_A - \text{h.c.}\bigr),\hat\rho\right] + \gamma\,\mathcal{D}\!\bigl[\hat\sigma^-_A + \hat\sigma^-_B\bigr]\hat\rho.
\label{eq:Rabi-clean}
\end{equation}
The antisymmetric coherent term inside the commutator encodes the unidirectional photon exchange characteristic of the cascade, and the collective jump operator $\hat\sigma^-_A + \hat\sigma^-_B$ reflects the indistinguishability of emissions from the two qubits into the shared waveguide. The unique pure dark state, written in the computational basis $\{\ket{g^c},\ket{e^c}\}$, is
\begin{equation}
\ket{\psi^c_{\rm ss}} = \frac{1}{\mathcal{N}}\!\left[\ket{g^c_A g^c_B} + \frac{2\Omega}{i\gamma}\bigl(\ket{e^c_A g^c_B}-\ket{g^c_A e^c_B}\bigr)\right],
\label{eq:Rabi-dark}
\end{equation}
which interpolates between $\ket{g^c_A g^c_B}$ at $\Omega/\gamma\to 0$ and the singlet at $\Omega/\gamma\to\infty$, with concurrence tunable by the single ratio $\Omega/\gamma$. Unlike the TMS case, \cref{eq:Rabi-dark} is not Schmidt-diagonal in the bare $\{\ket{g^c},\ket{e^c}\}$ basis. The Schmidt basis $\{\ket{0^c},\ket{1^c}\}$ used in \cref{sec:cycle-dynamics} is obtained by a local $\hat\sigma^x$ rotation
\begin{equation}
R(\theta) \equiv \exp\!\bigl(i\tfrac{\theta}{2}\hat\sigma^x_A\bigr)\otimes\exp\!\bigl(-i\tfrac{\theta}{2}\hat\sigma^x_B\bigr),\qquad \tan\theta = \frac{4\Omega}{\gamma},
\end{equation}
under which $R(\theta)\ket{\psi^c_{\rm ss}}$ takes the Schmidt form $u'\ket{0^c_A 0^c_B} + v'\ket{1^c_A 1^c_B}$ with $u' = \cos^2(\theta/2)/\mathcal{N}'$, $v' = - \sin^2(\theta/2)/\mathcal{N}'$, $\mathcal{N}' = \sqrt{(1+\cos^2\theta)/2}$. This Schmidt rotation is absorbed into a redefinition of the communication-qubit-controlled gates $\mathcal{C}_Z$ and $\mathcal{C}_X$ used in the cycle of \cref{sec:cycle-dynamics}, requiring only additional standard single-qubit control on the communication qubits. A similar argument applies to the protocol replacing the chiral waveguide with a bidirectional waveguide\cite{Shah-prxQuantum-2024-m}.

\subsection{Master equation with finite transmission loss}
\label{sec:impl-noisy}

In a realistic distributed setting the field linking the two nodes incurs propagation loss. We model the lossy link as a beam splitter with amplitude transmission $\eta\in[0,1]$, power transmission $|\eta|^2$, and power loss $\epsilon \equiv 1-|\eta|^2$. The corresponding open-system master equations follow from the SLH input-output framework \cite{Irfan-physRevRes-2024-v}: one composes the qubit-field SLH triples with the beam-splitter SLH triple and reads off the effective Lindbladian acting on the qubit reduced state.

For the TMS scheme, each output mode of the two-mode squeezed reservoir passes through a beam splitter of transmission $\eta$ before reaching its target qubit. The composition gives
\begin{align}
\mathcal{L}^{\rm TMS}_{\rm noisy}\hat\rho &= \gamma(1-\epsilon)\,\mathcal{D}\!\bigl[\cosh r\,\hat\sigma^-_A - \sinh r\,\hat\sigma^+_B\bigr]\hat\rho + 
\gamma(1-\epsilon)\,\mathcal{D}\!\bigl[\cosh r\,\hat\sigma^-_B - \sinh r\,\hat\sigma^+_A\bigr]\hat\rho\nonumber\\
&\quad + \gamma\epsilon\Bigl(\mathcal{D}[\cosh r\,\hat\sigma^-_A]+\mathcal{D}[\cosh r\,\hat\sigma^-_B]+\mathcal{D}[\sinh r\,\hat\sigma^+_A]+\mathcal{D}[\sinh r\,\hat\sigma^+_B]\Bigr)\hat\rho.
\label{eq:TMS-noisy}
\end{align}
The first line is the cooperative dissipator of \cref{eq:TMS-clean}, scaled by the power transmission $(1-\epsilon)$. The second line is the new contribution from the lost mode: a photon that escapes en route is no longer correlated with its partner, which results in the partner seeing the thermal bath with thermal occupation $n_{\rm th} = \sinh^2 r$. These local jumps do not share the dark state of \cref{eq:TMS-dark}; they pump the joint state out of the dark-state manifold at rate $\sim\gamma\epsilon\sinh^2 r$, which sets the leading-order leakage we analyze.

For the Rabi-driven cascade the field propagates strictly from $A$ to $B$, so loss enters asymmetrically. Emissions from $A$ are attenuated before reaching $B$; emissions originating at $B$ are unaffected by the link. The SLH composition then yields \cite{Irfan-physRevRes-2024-v}
\begin{align}
\mathcal{L}^{\rm Rabi}_{\rm noisy}\hat\rho &= -i\!\left[\Omega(\hat\sigma^x_A+\hat\sigma^x_B) - \tfrac{i\gamma\eta}{2}\bigl(\hat\sigma^+_B\hat\sigma^-_A - \text{h.c.}\bigr),\hat\rho\right]\nonumber\\
&\quad + \gamma\,\mathcal{D}\!\bigl[\eta\,\hat\sigma^-_A + \hat\sigma^-_B\bigr]\hat\rho + \gamma\epsilon\,\mathcal{D}[\hat\sigma^-_A]\hat\rho.
\label{eq:Rabi-noisy}
\end{align}
Three modifications relative to \cref{eq:Rabi-clean} are visible. (i) The cascade coherent exchange is rescaled by $\eta$, reflecting the reduced amplitude of $A$'s field reaching $B$. (ii) The collective jump operator becomes the asymmetric $\eta\,\hat\sigma^-_A + \hat\sigma^-_B$: $A$'s contribution is attenuated, while $B$'s remains at unit amplitude. (iii) A new local jump $\sqrt{\gamma\epsilon}\,\hat\sigma^-_A$ captures the fraction of $A$'s emission that is lost to the environment before reaching $B$.

% -------------------------------------------------------
\section{TMS Detection Probability: Analytic Derivation}
\label{sec:tms-pdet}
% -------------------------------------------------------

The conversion probability $p_{\rm conv}$ defined in \cref{eq:pdet-def} controls both the per-cycle convergence rate $\Gamma = -\ln(1-p_{\rm conv})$ and the optimal cycle time of the protocol. In this section we evaluate it analytically for the TMS dissipator of \cref{eq:TMS-clean}, valid at arbitrary squeezing $r$, via the quantum-trajectory unraveling of the master equation. The derivation collapses the continuous evolution into a discrete counting problem and yields a closed-form optimal verification time $\tau^*_{\rm conv}$.

\subsection{Physics of the unraveling}
\label{sec:unraveling-physics}

We want the probability that the communication qubits land in the odd-parity subspace at the end of the verification step $\mathcal{M}(\tau_{\rm conv})$, starting from the error state $\ket{\psi^c_{\rm det}} = u\ket{0^c_A 0^c_B} - v\ket{1^c_A 1^c_B}$ identified in \cref{eq:Psi2}. Direct integration of the master equation leads to additional complexity; quantum trajectory unraveling \cite{Wiseman-physRevA-1994-n} gives a much more transparent picture.

In the unraveling, the master-equation evolution is recast as the ensemble average over many pure-state \textit{trajectories}, each defined by a specific sequence of stochastic ``click'' times $\{t_1, t_2, \dots\}$ and click labels $\{a_1, a_2, \dots\}$. Between clicks the state evolves deterministically under the non-Hermitian Hamiltonian of TMS,
\begin{equation}
\hat H_{\rm nh} \;=-\frac{i\gamma}{2}\sum_{i}\hat L_i^\dagger \hat L_i
\end{equation}
which damps the norm at a rate set by the instantaneous emission probability. At each click time $t_i$, the jump operator $\hat L_{a_i}$ acts on the state and the wavefunction is renormalized. Tracing over the (otherwise unrecorded) click history reproduces the unconditional master equation.

The key fact that makes this unraveling tractable for our problem is a weak symmetry of the TMS Lindbladian: \emph{$\hat H_{\rm nh}$ conserves the joint communication parity} $\hat P_c \equiv \hat\sigma^z_A\hat\sigma^z_B$, while each jump $\hat L_i$ flips it. In any single trajectory the parity at the end of the verification step is therefore determined entirely by the parity of the click count along that trajectory. Since the initial state $\ket{\psi^c_{\rm det}}$ has even parity, the detection event ``communication qubits end in odd parity'' is precisely the event ``odd number of clicks occurred during $\tau_{\rm conv}$.'' We have reduced a continuous-time parity dynamics question to a discrete counting question:
\begin{equation}
p_{\rm conv}(\tau_{\rm conv}) \;=\; \mathrm{Prob}\bigl[\,\text{odd number of clicks in }\tau_{\rm conv}\;\big|\; \hat \rho (t=0) = \ket{\psi^c_{\rm det}} \bra{\psi^c_{\rm det}}\,\bigr].
\label{eq:pdet-odd-clicks}
\end{equation}

The weak $Z_2$ symmetry guarantees that parity is conserved by the deterministic no-jump dynamics ($[\hat H_{\rm nh},\hat P_c]=0$), so all parity changes during a trajectory are accounted for by the discrete click record. The Rabi-driven cascade of \cref{eq:Rabi-clean} does not enjoy even this weak symmetry: the local Rabi drive $\Omega(\hat\sigma^x_A + \hat\sigma^x_B)$ inside $\hat H_{\rm Rabi}$ rotates between parity sectors, $[\hat H_{\rm Rabi},\hat P_c]\neq 0$, and the no-jump propagator can flip parity \emph{without} a click. The clean reduction \cref{eq:pdet-odd-clicks} therefore fails for the Rabi protocol: $p_{\rm conv}$ as a function of $\tau_{\rm conv}$ acquires an intrinsically richer landscape with no closed-form trajectory sum, and must be evaluated by direct numerical integration of the full master equation --- as we do for the Rabi panels of Fig.~2 in the main text.

\subsection{Closed-form result and optimal $\tau_{\rm conv}$}
\label{sec:pdet-closed-form}

Two further structural features collapse the trajectory sum into an elementary expression. First, the no-jump dynamics inside each parity sector is essentially trivial: in the even sector $\hat H_{\rm nh}$ reduces to a $2\times 2$ block that maps $\ket{\psi^c_{\rm det}}$ back onto itself (up to a decaying norm), while in the odd sector it is proportional to the identity. Second, paired clicks of \emph{different} types ($\hat L_1$ followed by $\hat L_2$ or vice versa) return the state to a multiple of $\ket{\psi^c_{\rm det}}$, while same-type pairs annihilate against the initial state. The sum over trajectories with $2N+1$ clicks then collapses into a convolution that is geometric in Laplace space, and resuming over $N$ yields
\begin{equation}
{\;p_{\rm conv}(\tau_{\rm conv}) \;=\; \frac{16\,u^2 v^2}{\sqrt{1+32\,u^2 v^2}}\;e^{-\frac{3\gamma'\tau_{\rm conv}}{2}}\,\sinh\!\left(\frac{\sqrt{1+32\,u^2 v^2}}{2}\,\gamma'\tau_{\rm conv}\right),\;}
\label{eq:pdet-closed}
\end{equation}
where $\gamma' \equiv \gamma\cosh 2r$ is the effective dissipative rate set by the photon number of the squeezed bath and $u^2 v^2 = \tanh^2 2r/4$. This already exposes the qualitative trade-off: $p_{\rm conv}$ vanishes both at $\tau_{\rm conv}\to 0$ (no time for a photon to be emitted) and $\tau_{\rm conv}\to\infty$ (dynamics reaches the steady state and the error information is lost), and is maximized at an intermediate time.

Maximizing \cref{eq:pdet-closed} over $\tau_{\rm conv}$ yields the optimal conversion time in closed form,
\begin{equation}
\gamma'\,\tau^*_{\rm conv}(r) \;=\; \frac{1}{\alpha_r}\,\ln\!\left(\frac{3+\alpha_r}{3-\alpha_r}\right),\qquad \alpha_r \;\equiv\; \sqrt{1+8\tanh^2 2r},
\label{eq:tauver-opt}
\end{equation}
which interpolates between $\gamma\tau^*_{\rm conv}\to\ln 2$ at $r\to 0$ (recovering the single-jump-dominated result $\gamma\tau^*_{\rm conv}=\ln 2/\cosh 2r$) and a logarithmically growing optimum at large $r$. Substituting \cref{eq:tauver-opt} back into \cref{eq:pdet-closed} gives the maximum per-cycle detection probability $p^*_{\rm conv}(r)$ and hence the maximum stabilization rate $\Gamma_{\max}(r) = -\ln[1-p^*_{\rm conv}(r)]$, which is the dashed analytic curve in Fig.~2(a) and the dashed locus of optimal $\gamma\tau^*_{\rm conv}$ in Fig.~2(b) of the main text. The agreement with the full master-equation simulation across all $r$ confirms that the unraveling collapse to a click-counting problem captures the parity dynamics of the TMS verification step exactly. In contrast, the coherent-driven scheme does not have weak symmetry, therefore exhibits a more complex landscape of stabilization rate in \cref{fig:optimal-noise-params}(a).
 
% -------------------------------------------------------
\section{Saturation of Stabilization Rate with \texorpdfstring{$\tau_{\rm stb}$}{tau\_stb}}
\label{sec:saturation}
% -------------------------------------------------------

\begin{figure}[tb]
\centering
\includegraphics{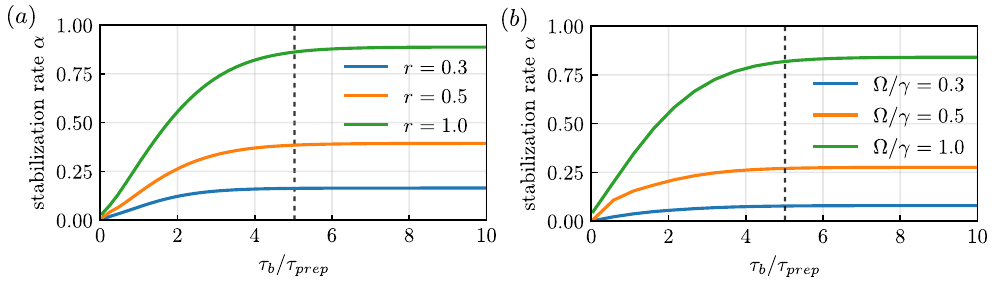}
\caption{Per-cycle stabilization rate $\Gamma$ as a function of $\tau_{\rm stb}/\tau_{\rm ent}$, with $\tau_{\rm conv}$ separately optimized at each point, for (a) TMS and (b) Rabi driven qubits schemes. The rate saturates within a few $\tau_{\rm ent}$; the dashed line marks the operating choice $\tau_{\rm stb} = 5\,\tau_{\rm ent}$ used throughout the main text.}
\label{fig:tau-stb-saturation}
\end{figure}

The analysis of \cref{sec:cycle-dynamics} treats the resource-preparation step as exact: $\mathcal{M}(\tau_{\rm stb})$ is assumed to drive the communication qubits all the way to the pure dark state $\ket{\psi^c_{\rm ss}}$ before parity encoding. We emphasize first that the fixed point of the cycle is independent of $\tau_{\rm stb}$: for \textit{any} $\tau_{\rm stb}>0$, the target product state $\ket{\Phi^L_+}\otimes\ket{\psi^c_{\rm ss}}$ remains a stationary point of $\mathcal{E}_{ZZ}$ and $\mathcal{E}_{XX}$, since $\ket{\psi^c_{\rm ss}}$ is the unique dark state of $\mathcal{M}(\tau_{\rm stb})$. The choice of $\tau_{\rm stb}$ therefore affects only the per-cycle convergence rate, not the steady state. At finite $\tau_{\rm stb}$, the input to $\mathcal{C}_Z$ retains an admixture of bright modes of $\mathcal{L}_{\rm ent}$, which (i) lowers the overlap between the error branch and $\ket{\psi^c_{\rm det}}$ in \cref{eq:Psi2} and (ii) seeds a small amount of odd-parity weight in the dark-state branch. Both effects decay exponentially with $\tau_{\rm stb}$ at a rate set by the dissipative (Lindblad) gap $\lambda_{\rm ent}$ of $\mathcal{L}_{\rm ent}$, so the natural reference timescale is the state-preparation time
\begin{equation}
\tau_{\rm ent} \equiv \lambda_{\rm ent}^{-1}.
\label{eq:tau-prep-def}
\end{equation}
Once $\tau_{\rm stb}\gtrsim$ a few $\tau_{\rm ent}$, the residual contamination is exponentially small and the per-cycle stabilization rate $\Gamma = -\ln(1-p_{\rm conv})$ saturates at its $\tau_{\rm stb}\to\infty$ value. Full simulations confirm that $\tau_{\rm stb}\gtrsim 5\,\tau_{\rm ent}$ suffices in both schemes (\cref{fig:tau-stb-saturation}); this is the convention we adopt throughout the main text.

The price for raising $\tau_{\rm stb}$ above this threshold is paid in wall-clock cycle time, so $\tau_{\rm ent}$ sets the minimum useful cycle duration. For the TMS dissipator of \cref{eq:TMS-clean}, the slowest relaxation mode is set by the cooperative jump operators \cite{Pocklington-physRevB-2022-p}. In the weak-squeezing regime the gap is essentially the bare emission rate, $\lambda^{\rm TMS}_{\rm gap}\sim\gamma\cosh 2r$, while in the strong-squeezing regime it closes inversely with the effective bath occupation,
\begin{equation}
\lambda^{\rm TMS}_{\rm gap} \;\sim\; \frac{\gamma}{\sinh^2 r}\qquad(r\gg 1),
\end{equation}
giving $\gamma\tau_{\rm ent}\sim\sinh^2 r$. For the Rabi-driven cascade of \cref{eq:Rabi-clean}, the dark state \cref{eq:Rabi-dark} approaches the singlet as $\Omega/\gamma\to\infty$, and the gap closes quadratically in the dimensionless drive,
\begin{equation}
\lambda^{\rm Rabi}_{\rm gap} \;\sim\; \frac{\gamma^3}{\Omega^2}\qquad(\Omega/\gamma\gg 1),
\end{equation}
giving $\gamma\tau_{\rm ent}\sim (\Omega/\gamma)^2$. In both cases the closing gap places a hard floor on the per-cycle wall-clock time once a significant entanglement resource is required, and underlies the existence of an optimal $\tau_{\rm stb}$ in the cat-qubit implementation [Fig.~6 of the End Matter], where the preparation cost competes against logical idling errors.

% -------------------------------------------------------
\section{Stabilization under finite transmission loss}
\label{sec:Noisy-master}
% -------------------------------------------------------

The noisy entangling Lindbladians \cref{eq:TMS-noisy,eq:Rabi-noisy} degrade the protocol through two distinct mechanisms. First, the steady state $\hat\rho^{\epsilon}_{\rm ss}\equiv\lim_{t\to\infty}e^{t\mathcal{L}^{\epsilon}_{\rm ent}}(\hat\rho_0)$ is no longer the pure dark state of \cref{sec:cycle-dynamics}; it acquires an odd-parity admixture that contaminates the resource-preparation step. Second, that same admixture causes the verification step to incorrectly activate even when the logical qubits carry the correct parities, producing a finite false-positive rate. We show below that both effects are linear in $\epsilon$ and yield a steady-state concurrence error $1-\mathcal{C}\propto\epsilon$. The derivation is carried out explicitly for the TMS scheme.  The Rabi case has the same structure but lacks the parity symmetry that makes a closed-form analysis tractable, hence we analyze it numerically. 

\subsection{Modified rate equation}
\label{sec:noisy-eom}

Repeating the joint-probability accounting of \cref{sec:ZZ-eom} with the noisy input $\hat\rho^{\epsilon}_{\rm ss}$ introduces a single new conditional probability, the loss-induced false-positive rate
\begin{equation}
p_{\rm err} \;\equiv\; p(P^c_{\rm odd}\,|\,P^L_{\rm even}) \;=\; \tr\!\left[\hat P^c_{\rm odd}\,\mathcal{M}(\tau_{\rm conv})(\hat\rho^{\epsilon}_{\rm ss})\right],
\label{eq:perr-def}
\end{equation}
which quantifies how often the verification triggers a logical flip when the logical parity was already correct. After one $Z^L Z^L$-correction cycle, the parity update of \cref{eq:ZZ-result} generalizes to
\begin{equation}
\langle\hat Z^L_A\hat Z^L_B\rangle'_n \;=\; (p_{\rm conv} - p_{\rm err}) \;+\; (1 - p_{\rm conv} - p_{\rm err})\,\langle\hat Z^L_A\hat Z^L_B\rangle_n,
\label{eq:ZZ-noisy}
\end{equation}
which reduces to \cref{eq:ZZ-result} at $p_{\rm err}=0$. The intervening $X^L X^L$-correction cycle additionally multiplies $\langle\hat Z^L_A\hat Z^L_B\rangle$ by the noisy communication-qubit correlator
\begin{equation}
\langle\hat\sigma^z_A\hat\sigma^z_B\rangle_{\rm ss} \;\equiv\; \tr_c\!\bigl(\hat\sigma^z_A\hat\sigma^z_B\,\hat\rho^{\epsilon}_{\rm ss}\bigr),
\label{eq:szz-ss-def}
\end{equation}
by the same Heisenberg-picture calculation that produced \cref{eq:XX-during-ZZ}. The full one-cycle update is therefore
\begin{equation}
\langle\hat Z^L_A\hat Z^L_B\rangle_{n+1} \;=\; \langle\hat\sigma^z_A\hat\sigma^z_B\rangle_{\rm ss}\,\bigl[(p_{\rm conv} - p_{\rm err}) + (1 - p_{\rm conv} - p_{\rm err})\,\langle\hat Z^L_A\hat Z^L_B\rangle_n\bigr],
\label{eq:ZZ-noisy-full}
\end{equation}
with an identical expression for $\langle\hat X^L_A\hat X^L_B\rangle$ by symmetry. The steady-state correlators are set by the competition between $p_{\rm conv}$, $p_{\rm err}$, and $1-\langle\hat\sigma^z_A\hat\sigma^z_B\rangle_{\rm ss}$, all of which we now show scale as $\mathcal{O}(\epsilon)$.

\subsection{Linear-in-\texorpdfstring{$\epsilon$}{epsilon} scaling}
\label{sec:perturb-rho}

To evaluate \cref{eq:ZZ-noisy-full} we need $\hat\rho^{\epsilon}_{\rm ss}$ to leading order in $\epsilon$. Treating the local relaxation terms in the second line of \cref{eq:TMS-noisy} as a perturbation of the cooperative dissipator (first line), Lindblad perturbation theory gives the ansatz
\begin{equation}
\hat\rho^{\epsilon}_{\rm ss} \;=\; \ket{\psi^c_{\rm ss}}\bra{\psi^c_{\rm ss}} \;+\; \epsilon\,\hat\rho_{1} \;+\; \mathcal{O}(\epsilon^2),
\label{eq:rho-eps}
\end{equation}
where $\hat\rho_{1}$ is a traceless first excited state of $\mathcal{L}_{\rm ent}$ and $c=\tr(\hat P_{\rm odd} \hat \rho_1)$ is its support on the odd-parity subspace which has no dependence on $\epsilon$. The structure of the perturbation ensures that the first-order correction will have finite overlap with the odd parity subspace: each local loss jump in \cref{eq:TMS-noisy} flips $\hat\sigma^z_A\hat\sigma^z_B$ exactly once, while the cooperative dissipator preserves parity, so the leading correction to the (even-parity) dark state must overlap with the odd-parity subspace.

Substituting \cref{eq:rho-eps} into \cref{eq:perr-def,eq:szz-ss-def}, and using the parity assignments $\hat\sigma^z_A\hat\sigma^z_B = +1$ in the even sector and $-1$ in the odd sector together with $\mathcal{M}(\tau_{\rm conv})\hat\rho^{\epsilon}_{\rm ss}=\hat\rho^{\epsilon}_{\rm ss}$, one immediately reads off
\begin{equation}
p_{\rm err} \;=\; c\,\epsilon + \mathcal{O}(\epsilon^2),\qquad \langle\hat\sigma^z_A\hat\sigma^z_B\rangle_{\rm ss} \;=\; 1 - 2c\,\epsilon + \mathcal{O}(\epsilon^2).
\label{eq:eps-corrections}
\end{equation}
Plugging these into \cref{eq:ZZ-noisy-full} and solving for the fixed point gives, to leading order in $\epsilon$,
\begin{equation}
\langle\hat Z^L_A\hat Z^L_B\rangle_{\rm ss} \;=\; \frac{p_{\rm conv} - c\,\epsilon}{p_{\rm conv} + 3c\,\epsilon} \;+\; \mathcal{O}(\epsilon^2) \;=\; 1 \;-\; \frac{4\,c\,\epsilon}{p_{\rm conv}} \;+\; \mathcal{O}(\epsilon^2),
\label{eq:ZZ-ss-noisy}
\end{equation}
and an identical result for $\langle\hat X^L_A\hat X^L_B\rangle_{\rm ss}$ by symmetry. Through the concurrence lower bound $\mathcal{C}\geq\langle\hat X^L_A\hat X^L_B\rangle + \langle\hat Z^L_A\hat Z^L_B\rangle - 1$, the steady-state concurrence error inherits the same scaling,
\begin{equation}
1 - \mathcal{C} \;\lesssim\; \frac{8\,c\,\epsilon}{p_{\rm conv}} \;+\; \mathcal{O}(\epsilon^2),
\label{eq:C-error-scaling}
\end{equation}
confirming the linear $\propto\epsilon$ behavior observed in Fig.~3 of the main text. We see that in the limit of $\epsilon = 0$, the concurrence is $1$ regardless of our choice of $r$.
This reflects the fact that we are always considering the steady state, i.e.~the state after many repetitions of the channel.  Without loss, we could either take $r$ very small and repeat the channel many many times, or take a larger $r$ and repeat the channel a fewer number of times.  

For finite $\epsilon$, the above degeneracy is lose.  The prefactor $c/p_{\rm conv}$ of the order-$\epsilon$ correction controls the amount of entanglement the steady state encodes. This motivates us for the following question: for a given weak $\epsilon$, how do we pick our entanglement parameter $r$, such that the steady state has the smallest concurrence error. Note that for this question, we always assume the system have already reached the steady state, i.e. we only study the steady state property $\hat \rho_{\rm ss} = \lim_{n \to \infty} \mathcal{E}^n (\hat \rho_0)$.

\subsection{Estimate of perturbation and the optimal operating point}
\label{sec:c-estimate}

First we use a heuristic expression as an estimate for the coefficient $c(r)$, as the ratio of the rate at which loss pushes population out of the dark-state manifold to the rate at which the cooperative dissipator returns it:
\begin{equation}
c \;\sim\; \frac{\sum_i \bigl\|\hat L^{\rm loss}_i\ket{\psi^c_{\rm ss}}\bigr\|^2/\gamma}{\lambda_{\rm ent}^{\rm TMS}/\gamma}.
\label{eq:c-estimate}
\end{equation}
This expression comes from the solution of a simple Markovian rate equation. The numerator follows from the four local jumps in the second line of \cref{eq:TMS-noisy}, which is the rate for the steady state leaves the even manifold; the denominator uses the weak-squeezing form $\lambda^{\rm TMS}_{\rm gap}\sim\gamma\cosh 2r$ from \cref{sec:saturation}, describing the rate coming back to the steady state manifold. This gives
\begin{equation}
c(r) \;\approx\; \frac{4\sinh^2 r\,\cosh^2 r}{\cosh^2 2r} \;=\; \tanh^2 2r,
\label{eq:c-tanh}
\end{equation}
which interpolates between $c\sim 4r^2$ at weak squeezing and $c\to 1$ at strong squeezing.

\begin{figure}[tb]
\centering
\includegraphics{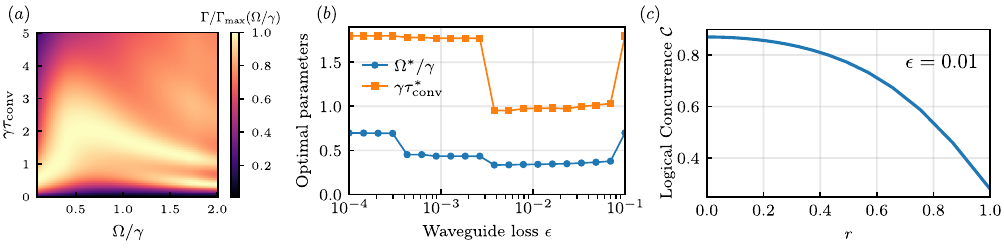}
% \fbox{\rule{0pt}{4cm}\rule{0.7\textwidth}{0pt}}
\caption{(a) Normalized stabilization rate $\Gamma/\Gamma_{\rm max}$ of coherent-driven scheme in the noiseless limit, mimicking Fig.~2(b) in the main text. (b) Optimal stabilization parameters as a function of waveguide loss $\epsilon$ used in Fig.~3. Here we show the optimal conversion time $\gamma\tau^*_{\rm conv}$ and optimal entanglement parameter $\Omega^*/\gamma$. (c) Logical steady state entanglement $\mathcal{C}$ as a function of the squeezing parameter $r$ of the TMS. The waveguide loss here is $\epsilon = 10^{-2}$. }
\label{fig:optimal-noise-params}
\end{figure}

Using this estimate, we next investigate what the optimal choice of $r$ is given the presence of weak loss.  The two factors in \cref{eq:C-error-scaling}, namely $c$ and $1/p_{\rm conv}$, compete as the entangling-resource parameter $r$ is tuned. Increasing $r$ raises $p_{\rm conv}$ \cref{eq:pdet-closed} but simultaneously raises $c$, and additionally closes the dissipative gap. Joint optimization of $(r,\tau_{\rm conv})$ at fixed $\epsilon$ produces the linear-scaling curves of Fig.~3 of the main text and yields an $\epsilon$-dependent optimal operating point. In numerical optimization, we found that the optimal operating point is achieved at asymptotic limit of $r\to 0$ [see \cref{fig:optimal-noise-params}(c). This comes from the fact that in the weak-squeezing limit the detection rate also scales as $p^*_{\rm conv}(r) = \frac{1}{2}\tanh^2 2r + \mathcal{O}(r^4)$, the same quadratic order as $c(r) = \tanh^2 2r \approx 4r^2$. However, the leading order correction to $p^*_{\rm conv}(r)$ coefficient $32-48\ln 2\approx -1.27$ is negative. The concurrence error floor of \cref{eq:C-error-scaling} therefore reaches its smallest value, $1-\mathcal{C}\lesssim 16\,\epsilon$, at vanishing squeezing. This argument matches our numerical simulation of the noisy steady state entanglement in \cref{fig:optimal-noise-params}(c). Although $r\to 0$ serves as the optimal point for this case with waveguide loss only, and realistic noise model including finite per cycle loss, e.g. communication qubit or logical qubit error, will kill this advantage, as this is also the limit where the stabilization rate vanishes $\Gamma \to 0$. But it also gives us a nice surprise: in this case with only waveguide loss, we actually benefits from using a weaker entangling bath and concentrates slower to the logical qubit.

% To avoid confusion, we wish to stress that there is nothing contradictory about finding that $r \rightarrow 0$ yields optimal performance for weak loss.  We are looking at the parametric dependence of the steady state, hence we are implicitly {\it first} taking the number of channel uses to infinity before taking any parameter limits like $r \rightarrow 0$.  Hence, the results for $r=0$ should be interpreted as:
% % To correctly interpret this limit, we wish to point out the optimal at vanishing $r$ is under a specific order of limit. Our discussion about the steady state throughout the paper always take the infinite time limit first, namely
% \begin{equation}
%     \hat \rho_{ss}(r = 0) = \lim_{r \to 0} \lim_{n \to +\infty} \mathcal{E}^n (\hat \rho_0).
% \end{equation}
% Since the gap closes in $r \to 0$ limit, these two limits do not commute.

% -------------------------------------------------------
\section{Cat circuit dynamics}
\label{sec:cat-circuit}
% -------------------------------------------------------
In this section, we walk through the alternative circuit designed for cat qubits, using gate set $\{ \mathcal{C}_X, S\}$. The composite channel for each cycle is
\begin{equation}
\mathcal{E}_{\rm cat} \;=\; \mathcal{C}_X\,\circ\,\bigl[\mathbb{I}_L\otimes\mathcal{M}(\tau_{\rm conv})\bigr]\,\circ\,[\hat S_A \otimes \hat S_B \otimes \mathbb{I}_c]\,\circ\,\mathcal{C}_X\,\circ\,\bigl[\mathbb{I}_L\otimes\mathcal{M}(\tau_{\rm stb})\bigr],.
\label{eq:E-cat}
\end{equation}
At first glance, this does not match our protocol introduced first in the main text. To give a heuristic understanding of how this "reduced" circuit can stabilize both correlation of interest $\left\{ X^L_A X^L_B, Y^L_A Y^L_B \right \}$. Using $S X S^\dagger = Y$, we can rewrite $\mathcal{E}_{\rm cat}$ in two equivalent ways:
\begin{align}
    \mathcal{E}_{\rm cat} \;=\; \mathcal{C}_X\,\circ\,\bigl[\mathbb{I}_L\otimes\mathcal{M}(\tau_{\rm conv})\bigr]\,\circ\,\mathcal{C}_Y\,\circ\,\bigl[\mathbb{I}_L\otimes\mathcal{M}(\tau_{\rm stb})\bigr]\,\circ\,[\hat S_A \otimes \hat S_B \otimes \mathbb{I}_c], \\
    \mathcal{E}_{\rm cat} \;=\;[\hat S_A \otimes \hat S_B \otimes \mathbb{I}_c] \,\circ\,\mathcal{C}_{-Y}\,\circ\,\bigl[\mathbb{I}_L\otimes\mathcal{M}(\tau_{\rm conv})\bigr]\,\circ\,\mathcal{C}_X\,\circ\,\bigl[\mathbb{I}_L\otimes\mathcal{M}(\tau_{\rm stb})\bigr]. \label{eq-app:cat_channel_decomp}
\end{align}
We see that this mimic our previous subroutine up to a initial (final) phase gate. Combining them together, using $\hat S^2 = \hat Z$, we have
\begin{equation}
    \mathcal{E}_{\rm cat}^2 = \mathcal{E}_{YY} \, \circ \, [\hat Z_A \otimes \hat Z_B\otimes \mathbb{I}_c] \, \circ \, \mathcal{E}_{XX}.
\end{equation}
This is the connection we made in the main text. Here $\mathcal{E}_{YY}$ is identical to the main protocol, yet $\mathcal{E}_{XX}$ is measuring $XX$ yet feedback on $\mathcal{C}_{-Y}$. Since the feedback operation still satisfy $\{-Y_{A(B)},X_AX_B\}=0$ and $[-Y_{A(B)},Y_AY_B]=0$, we can confirm it performs exact same feedback effect as $\mathcal{C}_Y$.

\subsection{Operator dynamics}
Next, we investigate the operator dynamics of $XX$ and $YY$ for $\mathcal{E}_{\rm cat}$ in the limit of $\tau_{\rm stb} \gg \tau_{\rm ent}$. For each application of $\mathcal{E}_{\rm cat}$, using~\cref{eq-app:cat_channel_decomp}, we have
\begin{align}
    \langle \hat X_A^L \hat X_B^L \rangle_{n+1} &= \langle \hat \sigma_A^Z \hat \sigma_B^Z \rangle_{\rm ss}\langle \hat Y_A^L \hat Y_B^L \rangle_{n},\\
    \langle \hat Y_A^L \hat Y_B^L \rangle_{n+1} &= \langle \hat X_A^L \hat X_B^L \rangle_{n} + p_{\rm conv} (1- \langle \hat X_A^L \hat X_B^L \rangle_{n}).
\end{align}
We can read this directly from~\cref{eq-app:cat_channel_decomp}: it works as the $\mathcal{E}_{\rm cat} = [\hat S_A \otimes \hat S_B \otimes \mathbb{I}_c] \, \circ \, \mathcal{E}_{\rm XX}$, a XX parity concentration followed by a basis change between X and Y. This also reflects in the operator dynamics above: $YY$ is concentrating and $XX$ is behind $YY$ by one step. One can see this in the Fig.~7(b) in the End Matter, that $YY$ is always leading $XX$ and they eventually converges to similar value in the steady state.

% -------------------------------------------------------
\section{Gate Realization for Different Logical Encoding}
\label{sec:gate-realization}
% -------------------------------------------------------

In this section, we discuss the experimental implementation of the stabilization protocol. The protocol places single operational requirements on the logical system at each node. There must exist a physical interaction between the communication qubit and the logical system from which a \textit{communication-qubit-controlled logical gate} can be engineered: specifically, a controlled-$Z_L$ (controlled-$X_L$) gate that applies the logical Pauli $Z_L$ ($X_L$) to the code space conditional on the communication qubit state, while leaving the code stabilizers intact. Crucially, neither requirement involves coupling the logical qubit directly to the entangling bath $\mathcal{L}_{\rm ent}$; the bath acts exclusively on the communication qubits, so any encoding that provides the above gate set while preserving the codespace under the local noise model is in principle eligible.

We discuss the gate realization for three paradigmatic encodings: cat qubits (Sec.~\ref{sec:gate-cat}), GKP qubits (Sec.~\ref{sec:gate-gkp}), and surface codes (Sec.~\ref{sec:gate-surface}).

\subsection{Cat qubits}
\label{sec:gate-cat}

For a cat qubit with logical basis $\ket{0^L} = \ket{+\alpha} + \mathcal{O}(e^{-|\alpha|^2})$ and $\ket{1^L} = \ket{-\alpha} + \mathcal{O}(e^{-|\alpha|^2})$, the photon-number parity operator $\hat P = e^{i\pi\hat a^\dagger\hat a}$ satisfies $\hat P\ket{\pm\alpha} = \ket{\mp\alpha}$, identifying $\hat P = X_L$ in the logical basis. Consequently, a transmon-controlled parity gate directly implements the required controlled-$X_L$ operation. We replace the $\{CX_L, CZ_L\}$ gate set of the general protocol with $\{CX_L, \hat S\}$, where $\hat S$ is a single-photon-driven cat phase gate. This is equivalent to stabilizing the Bell state with $\langle X_A^L X_B^L\rangle = 1$ and $\langle Y_A^L Y_B^L\rangle = 1$ (related to the $XX$/$ZZ$ Bell state by a local $S$ rotation), with both cycles realized entirely through the dispersive interaction.

The controlled-$X_L$ gate is implemented via the $\chi$-matched dispersive interaction $\hat H_{\rm dis} = \chi\hat a^\dagger\hat a\otimes(\ket{e}\bra{e} + \ket{f}\bra{f})$ introduced in Ref.~\cite{Guillaud-physRevX-2019-a}. Free evolution for time $t_g = \pi/\chi$ generates
\begin{equation}
  U(t_f) = \hat{\mathbb{I}}\otimes\ket{g}\bra{g} + \hat P\otimes\left(\ket{e}\bra{e}+\ket{f}\bra{f}\right),
  \label{eq:CX-gate}
\end{equation}
which applies $X_L = \hat P$ to the cat qubit conditional on the transmon being in $\ket{e}$ or $\ket{f}$, and leaves it unchanged when the transmon is in $\ket{g}$. The $\ket{f}$ level is used to ensure the gate remains unaffected by transmon decay $\ket{e}\to\ket{g}$ during the operation: a decay from $\ket{f}\to\ket{e}$ produces the same logical action as $\ket{e}$, preserving the controlled operation.

The gate in \cref{eq:CX-gate} is bias-preserving. A photon-loss event on the cat cavity at time $t_1$ within the gate produces the error operator
\begin{equation}
  \hat K_{t_1} \propto \hat U(t_g - t_1)\,\hat a\,\hat U(t_1) = \hat a\, e^{-i\chi(t_g-t_1)(\ket{e}\bra{e}+\ket{f}\bra{f})}\,\hat U(t_g).
\end{equation}
Since $\hat a\ket{\pm\alpha} = \pm\alpha\ket{\pm\alpha}$, the $\hat a$ factor acts as $Z_L$ on the cat logical subspace. The remaining phase factor $e^{-i\chi(t_g-t_1)(\ket{e}\bra{e}+\ket{f}\bra{f})}$ is a $Z$-rotation on the transmon. Integrating over all loss times, the error channel decomposes into Kraus operators
\begin{equation}
  \hat K_\pm \propto Z_c\otimes e^{\pm i\frac{\pi}{4}\hat\sigma^z},
\end{equation}
where $Z_c = \hat a / |\alpha|$ denotes the logical $Z_L$ on the cat. All error events are thus $Z$-type on both the cat qubit and the transmon, and no $X_L$ errors are generated. The exponential suppression of cat bit-flip errors ($X_L$ errors) with $|\alpha|^2$ is therefore preserved throughout the gate.

\subsection{GKP qubits}
\label{sec:gate-gkp}

For an ideal (infinite-energy) GKP qubit \cite{Gottesman-physRevA-2001-d}, the logical Pauli operators are displacement operators in orthogonal quadratures of the oscillator:
\begin{equation}
  Z_L = e^{i\sqrt{\pi}\hat{q}} = D(i\sqrt{\pi/2}), \qquad X_L = e^{-i\sqrt{\pi}\hat{p}} = D(\sqrt{\pi/2}),
\end{equation}
where $\hat{q} = (\hat{a}+\hat{a}^\dagger)/\sqrt{2}$, $\hat{p} = i(\hat{a}^\dagger-\hat{a})/\sqrt{2}$, and $D(\alpha) = e^{\alpha\hat{a}^\dagger - \alpha^*\hat{a}}$ is the displacement operator. The codespace stabilizers are $S_q = Z_L^2 = D(i\sqrt{2\pi})$ and $S_p = X_L^2 = D(\sqrt{2\pi})$. The required controlled-$Z_L$ and controlled-$X_L$ gates therefore both reduce to a \textit{conditional displacement} of the GKP cavity conditional on the transmon state. Concretely, defining $CD(\beta) = D(\beta)|g\rangle\langle g| + D(-\beta)|e\rangle\langle e|$, the two gates correspond to $CD(i\sqrt{\pi/2})$ and $CD(\sqrt{\pi/2})$ respectively --- the same gate type, differing only in the quadrature direction of the displacement.

Conditional displacements are experimentally accessible in superconducting circuits via the echoed conditional displacement (ECD) technique \cite{CampagneIbarcq-nature-2020}, which implements $CD(\beta)$ using a sequence of transmon $\pi$-pulses interleaved with cavity drives under the dispersive interaction $\hat H_{\rm dis} = -\chi|e\rangle\langle e|\hat{a}^\dagger\hat{a}$. The ECD protocol suppresses spurious photon-number-dependent phases and requires only standard qubit control and cavity drives already present in GKP circuit experiments \cite{Grimsmo-arxivquant-ph-2021-o}.

For a physical (finite-energy) GKP qubit, the logical states carry a Gaussian envelope with squeezing parameter $\Delta$, making them only approximate eigenstates of the stabilizers. The conditional displacement gates still implement the correct logical action on the codespace, but each gate introduces displacement errors of order $O(\Delta^2)$ that shift the oscillator state slightly outside the code lattice. Uncorrected, these errors --- compounded by photon loss, which generates small random displacements $\hat{a} \propto \hat{q}+i\hat{p}$ during idling --- degrade the logical fidelity over many stabilization cycles. The natural remedy is to interleave local GKP syndrome extraction between entanglement cycles: periodic measurement of $S_q$ and $S_p$ (e.g.\ via a SBS protocol or autonomous stabilization \cite{Grimsmo-arxivquant-ph-2021-o,Royer-physRevLett-2020-t}) restores the GKP state to the codespace using only local operations, in direct analogy with the role played by $\mathcal{L}_{\rm cat}$ in the cat qubit implementation. Since syndrome extraction is local and does not involve the shared bath $\mathcal{L}_{\rm ent}$, the structure of the entanglement stabilization protocol is fully preserved. One practical consideration is that the transmon communication qubit and the GKP syndrome ancilla are the same physical device in the most compact architecture; in this case, the transmon must be time-multiplexed between GKP stabilization and communication duties, which sets a lower bound on the entanglement cycle time.

\subsection{Surface codes}
\label{sec:gate-surface}

In a distance-$d$ surface code \cite{Kitaev-annPhysnY-2003-e,Dennis-jMathPhys-2002-z, Fowler-physRevA-2012-q}, the logical Pauli operators are transversal: $Z_L = \bigotimes_{i\in\mathcal{P}_Z} Z_i$ and $X_L = \bigotimes_{i\in\mathcal{P}_X} X_i$, where $\mathcal{P}_Z$ and $\mathcal{P}_X$ are paths of $d$ physical qubits connecting opposite boundaries of the code patch. Since these logical operators are products of single-qubit Paulis, the required controlled-$Z_L$ and controlled-$X_L$ gates decompose exactly into products of two-qubit gates between the communication transmon and the physical data qubits along each path:
\begin{equation}
  C Z_L = \prod_{i\in\mathcal{P}_Z} CZ_{\rm transmon,\,i}, \qquad CX_L = \prod_{i\in\mathcal{P}_X} CNOT_{\rm transmon \to i}.
\end{equation}
This follows from the fact that single-qubit $Z_i$ operators along $\mathcal{P}_Z$ mutually commute, so the controlled-$Z_L$ factorizes into a sequence of $d$ independent CZ gates. Each CZ (or CNOT) gate acts between the transmon and a single data qubit, uses only standard two-qubit interactions, and leaves all code stabilizers intact: the $Z$-type stabilizers (products of $Z$ on weight-4 plaquettes) commute with the additional $Z_i$s applied, and the $X$-type stabilizers have even overlap with any string operator and are therefore also preserved. The gate count scales as $O(d)$ per controlled logical operation.

Local error correction of the surface code --- periodic extraction of $X$- and $Z$-type stabilizer syndromes and classical decoding --- proceeds independently of the communication qubit and the shared bath $\mathcal{L}_{\rm ent}$, exactly as in a standalone surface code. The logical error rate per syndrome cycle scales as $(p/p_{\rm th})^{\lfloor(d+1)/2\rfloor}$, where $p$ is the physical error rate and $p_{\rm th}\approx 1\%$ is the surface code threshold \cite{Fowler-physRevA-2012-q}. By choosing $d$ sufficiently large, the logical error rate can be made negligible compared to $1/p_{\rm conv}$ (the number of stabilization cycles to reach steady state), ensuring that the logical qubit remains well-protected throughout the entanglement stabilization protocol.

\bibliography{ref}